\documentclass[prx,reprint]{revtex4-2}
\usepackage[dvips]{graphicx}
\usepackage{type1cm}
\usepackage[usenames,svgnames]{xcolor}
\usepackage{color}
\usepackage{url}
\usepackage{natbib}
\usepackage[colorlinks=true,linkcolor=blue,citecolor=blue,breaklinks=true]{hyperref}

\usepackage{amsmath,amssymb,bm,amsthm,mathtools}

\DeclareMathOperator{\Tr}{\mathrm{Tr}}

\newcommand{\pd}{\partial}

\newcommand{\ddv}[2]{\frac{d#1}{d#2}}

\usepackage{braket}

\usepackage[capitalize]{cleveref}
\crefname{section}{Sec.}{Secs.}
\Crefname{section}{Section}{Sections}

\newtheorem*{statement}{Accelerated dissipation}

\newcommand{\HT}{\hat{H}_\mathrm{T}}
\newcommand{\HS}{\hat{H}_\mathrm{S}}
\newcommand{\HB}{\hat{H}_\mathrm{B}}
\newcommand{\HI}{\hat{H}_\mathrm{I}}
\newcommand{\TrS}{\Tr_\mathrm{S}}
\newcommand{\TrB}{\Tr_\mathrm{B}}
\newcommand{\rhoT}{\rho_\mathrm{T}}
\newcommand{\rhoB}{\rho_\mathrm{B}}
\newcommand{\tauB}{\tau_\mathrm{B}}
\newcommand{\tildetauB}{\tilde{\tau}_\mathrm{B}}
\newcommand{\LI}{\mathcal{L}_\mathrm{I}}
\newcommand{\LS}{\mathcal{L}_\mathrm{S}}

\begin{document}
\nocite{apsrev41control}

\title{Error bounds on the universal Lindblad equation in the thermodynamic limit}

\author{Teruhiro Ikeuchi}
\author{Takashi Mori}
\affiliation{
Department of Physics, Keio University, Hiyoshi, Yokohama, 223-8522, Japan
}

\begin{abstract}
It is a central problem in various fields of physics to elucidate the behavior of quantum many-body systems subjected to bulk dissipation.
In this context, several microscopic derivations of the Lindblad quantum master equation for many-body systems have been proposed so far.
Among them, the universal Lindblad equation derived by Nathan and Rudner is fascinating because it has the desired locality and its derivation seems to rely solely on the assumption that the bath correlation time is much shorter than the dissipation time, which is the case in the weak-coupling limit or the singular-coupling limit.
However, it remains elusive whether errors from several approximations in deriving the universal Lindblad equation remain small during the time evolution in the thermodynamic limit.
Here, rigorous error bounds on the time evolution of a local quantity are given, and it is shown that, under the assumption of accelerated dissipation in bulk-dissipated systems, those errors vanish in the weak-coupling limit or the singular-coupling limit after taking the thermodynamic limit.
\end{abstract}
\maketitle


\section{Introduction}
The quantum Markov process with complete positivity is generally described by a quantum master equation of the Lindblad form~\citep{Lindblad1976,Gorini1976}.
The Lindblad equation has widely been used to describe a small quantum system in contact with thermal reservoirs, especially in quantum optics~\citep{Carmichael_text,Louisell_text}.
Recent interest has shifted to open quantum \emph{many-body} systems in connection with quantum control utilizing dissipation~\citep{Bloch2012,Barreiro2011,Barontini2013,Tomita2017}, noisy quantum computation~\citep{Bharti2022,Cai2023}, and nonequilibrium statistical physics~\citep{Diehl2010,Minganti2018,Cai2013,Bouganne2020,Rakovszky2024}.
The Lindblad equation is also a fundamental tool to study such a many-body system.
Recent theoretical studies on open quantum many-body systems are largely triggered by experimental progress, which enables us to introduce tunable dissipation in many-body setups~\citep{Barreiro2011,Bloch2012,Barontini2013,Tomita2017}.
In addition to this, there is also a line of research based on the idea that weak dissipation can be utilized as a probe field to better understand a given quantum many-body system~\citep{Pan2020,Mori2024_Liouvillian}.

New insights have been gained through recent theoretical studies on the many-body Lindblad dynamics.
In particular, it is recognized that operator scrambling dynamics~\citep{Nahum2018,Keyserlingk2018} has important implications for dissipative dynamics under bulk dissipation.
The operator scrambling---or operator spreading---implies that a local operator becomes highly nonlocal after quantum time evolution in the Heisenberg picture.
In general, a nonlocal operator is more fragile to dissipation, and hence the dissipative decay is accelerated because of the operator growth~\citep{Shirai2024_accelerated}, which was numerically confirmed in the decay of the Loschmidt echo fidelity that is faster than exponential~\citep{Schuster2023}.
As a consequence of such amplification of the effect of bulk dissipation, we sometimes have a non-zero decay rate even in the dissipationless limit (the thermodynamic limit is taken first), which is known as quantum Ruelle-Pollicott resonance~\citep{Mori2024_Liouvillian,Prosen2002}, or anomalous relaxation~\citep{Garcia-Garcia2023,Yoshimura2024}. 

In this way, fundamental properties of many-body Lindblad equations have been investigated, but their theoretical grounds are not well established.
It is highly desirable to give a microscopic derivation of the Lindblad equation and understand under what conditions its use is justified.
A standard derivation of the Lindblad equation involves using the secular approximation~\citep{Breuer_text} (it is also referred to as the rotating-wave approximation), and the derived equation is known as the Davies equation~\citep{Davies1974}.
The Davies equation is rigorously derived from the unitary time evolution of the system and the environment when the system of interest is finite and the system-bath interaction is infinitesimally weak.
However, it is later recognized that the secular approximation is not justified and gives unphysical results in the many-body setup~\citep{Wichterich2007,Mori2023_review}.
We have to go beyond the secular approximation to derive the Lindblad equation for open quantum many-body systems.

So far, based on different approximations, various derivations of the Lindblad equation for many-body systems have been proposed~\citep{Vacchini2000,Schaller2008,Wichterich2007,Nathan2020,Nathan2024,Gneiting2020,Becker2021,Shiraishi2024}.
Among them, the universal Lindblad equation (ULE) derived by \citet{Nathan2020,Nathan2024} has remarkable feature.
Firstly, it has desired locality~\citep{Nathan2020,Shiraishi2024} in contrast to the Davies equation.
Secondly, the ULE appears to be solely derived from the assumption that the bath correlation time is much shorter than the dissipation time, which is the case in the weak-coupling limit (the system-bath coupling is infinitesimally weak)~\citep{van_Hove1954,Davies1974} or in the singular-coupling limit (the bath correlation time is infinitesimally short)~\citep{Hepp1973,Gorini1976,Palmer1977}. 
No assumption on the timescale of the intrinsic dynamics of the system of interest is explicitly used.
Thus, it is expected that the derivation can safely apply to many-body systems.

Although the existing derivation of the ULE given in Ref.~\citep{Nathan2020} is physically reasonable, it is not rigorous and it is uncertain whether errors induced by several approximations needed to derive the ULE are negligible (see Ref.~\citep{TelloBreuer2024} for a numerical study on this problem).
\citet{Nathan2020} evaluated rigorous error bounds on the approximations, but there are two problems:
\begin{itemize}
\item[(i)] Their error bounds diverge in the thermodynamic limit for bulk dissipation (i.e. each site of the system is coupled to its own bath).
Therefore, it remains unclear whether those approximations are justified \emph{in macroscopic systems}.
\item[(ii)] They evaluated the error \emph{at the level of the equations of motion}.
It is not proved that the error does not accumulate over time.
Thus, it is uncertain whether the error remains small during the time evolution.
\end{itemize}

The aim of this paper is to overcome these two difficulties.
We consider a one-dimensional quantum spin system with local Hamiltonian in which each site is coupled to its own free-boson bath, and evaluate rigorous error bounds on the time evolution of a local quantity because of the approximations made in the derivation of the ULE.
Under the assumption of the \emph{accelerated dissipation}, which is expected to be a generic property of bulk-dissipated open quantum many-body systems~\citep{Schuster2023,Shirai2024_accelerated,Mori2024_Liouvillian}, we show that the errors vanish \emph{uniformly in time} when we take the weak-coupling limit or the singular-coupling limit \emph{after taking the thermodynamic limit}.
The ULE is therefore justified for arbitrarily long times in the thermodynamic limit.

To overcome the difficulty (i), we prove that the existing Lieb-Robinson bound~\citep{Nachtergaele2011} is applicable to the ULE.
By using it, finite error bounds in the thermodynamic limit are obtained.
To overcome the difficulty (ii), i.e. to obtain the desired time-independent error bounds, we use the assumption of accelerated dissipation.
Roughly speaking, it states that the dissipative decay rate of a local operator in the Heisenberg picture increases with time and is eventually saturated at a value that is much larger than the initial decay rate.
As we have already mentioned, it was argued that such an accelerated dissipation generically occurs because of the operator spreading~\citep{Schuster2023,Shirai2024_accelerated,Mori2024_Liouvillian}.
This is not a mathematically proven fact, but should be regarded as a reasonable conjecture.
We give its theoretical picture and numerical evidence under a local Lindblad dynamics.

The plan of the paper is as follows.
\Cref{sec:setup} introduces the microscopic setup.
\Cref{sec:main} gives a summary of the main result presented in this paper.
\Cref{sec:method} introduces theoretical methods used in deriving the main result: the Lieb-Robinson bound and accelerated dissipation.
The former is formulated as a rigorous inequality as is explained in \cref{sec:LR}, whereas the latter is a conjecture without mathematical proof.
\Cref{sec:acc} is devoted to a detailed account of accelerated dissipation with an intuitive picture and numerical evidence.
\Cref{sec:error} derives the error bound on the ULE by using theoretical methods.
\Cref{sec:derivation} reviews the microscopic derivation of the ULE following \citet{Nathan2020}.
Various approximations are introduced and the explicit expressions of their errors are given.
\Cref{sec:bound} illustrates how to use the Lieb-Robinson bound and the accelerated dissipation in evaluating the error induced by the Born approximation.
Full details of the derivation are given in \cref{sec:detail}.
We conclude with a discussion in \cref{sec:conclusion}.

\section{Summary of the main result}
\label{sec:summary}

Before going on to the theoretical detail, we summarize the main results presented in this paper.
We first explain the theoretical setup and then our main results and key theoretical methods.

\subsection{Setup}
\label{sec:setup}

Let us consider a one-dimensional quantum spin system defined on the set of sites $\Lambda=\{1,2,\dots,N\}$.
The results presented in this paper are straightforwardly extended to a $d$-dimensional system.
The Hamiltonian of the system of interest is denoted by $\HS$.
We assume that $\HS$ is local in the sense that it is written as
\begin{equation}
\HS=\sum_{i\in\Lambda}\hat{h}_i,
\end{equation}
where $\hat{h}_i$ is an operator acting nontrivially on sites near $i$.
For simplicity, we assume that $\{\hat{h}_i\}_{i\in\Lambda}$ are finite range: the diameter of the support of $\hat{h}_i$ is not greater than a constant $a$ for any $i\in\Lambda$.

In this paper, we focus on a static Hamiltonian, but the following discussion can be extended to a time-dependent Hamiltonian $\HS(t)$ by making a slight modification.

\begin{figure}[t]
\centering
\includegraphics[width=0.9\linewidth]{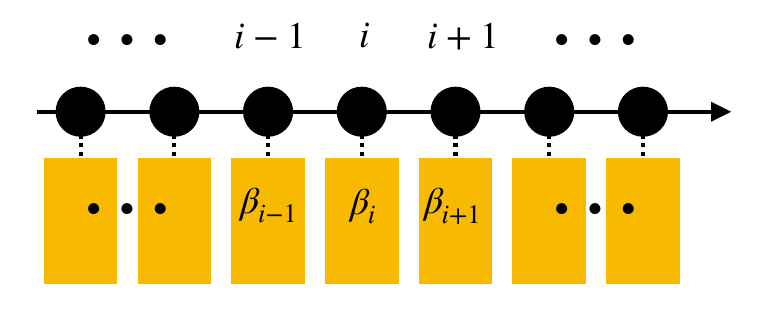}
\caption{Setup of a one-dimensional quantum many-body system under bulk dissipation.
Each site $i$ is coupled to an independent free-boson bath at the inverse temperature $\beta_i$.}
\label{fig:setup}
\end{figure}

We consider bulk dissipation, see \cref{fig:setup} for a schematic picture.
Each site $i$ is coupled to its own bath consisting of free bosons, and hence the Hamiltonian of the baths is written as
\begin{equation}
\HB=\sum_{i\in\Lambda}\HB^{(i)}, \quad \HB^{(i)}=\sum_k\omega_k\hat{b}_{i,k}^\dagger\hat{b}_{i,k},
\end{equation} 
where $\hat{b}_{i,k}^\dagger$ ($\hat{b}_{i,k}$) is the creation (annihilation) operator of bosons interacting with the site $i$.
The interaction between the system and the baths is expressed by
\begin{equation}
\HI=\sum_{i\in\Lambda}\hat{A}_i\otimes\hat{B}_i,
\end{equation}
where 
\begin{equation}
\hat{B}_i=\sum_k\left(g_k\hat{b}_{i,k}+g_k^*\hat{b}_{i,k}^\dagger\right)
\end{equation}
is the bath operator, which is assumed to be linear in $\hat{b}_{i,k}$ and $\hat{b}_{i,k}^\dagger$,
and $\hat{A}_i$ is a Hermitian operator of the system that is normalized as $\|\hat{A}_i\|=1$ ($\|\cdot\|$ is the operator norm).
For simplicity, in the following, we assume that $\hat{A}_i$ is a single-site operator, but it is straightforward to extend the following results to an arbitrary local operator $\hat{A}_i$ that acts nontrivially onto sites near $i$.
It is also straightforward to consider multiple dissipation channels at each site $i$: $\HI=\sum_{i\in\Lambda}\sum_{m=1}^M\hat{A}_{i,m}\otimes\hat{B}_{i,m}$ as long as $M$ is finite.

One might wish to consider a more general situation in which the bulk of the system is coupled to a single bath.
Such a situation is implemented by adding the interaction Hamiltonian $\hat{H}_\mathrm{B}^{(i,j)}$ between baths at different sites $i$ and $j$.
As long as $\hat{H}_\mathrm{B}^{(i,j)}$ is local in the sense that it decays sufficiently quickly as the distance between $i$ and $j$ increases, we expect that the results in this paper also apply to this case, but we do not enter this technical discussion.

The bath correlation function in the state 
\begin{equation}\label{eq:rhoB}
\rhoB=\frac{\exp\left[-\sum_{i\in\Lambda}\beta_i\HB^{(i)}\right]}{\TrB\exp\left[-\sum_{i\in\Lambda}\beta_i\HB^{(i)}\right]},
\end{equation} 
where $\beta_i$ is the inverse temperature of the $i$th bath, is defined by
\begin{equation}
\Phi_i(t)=\Tr_\mathrm{B}\left[\hat{B}_i(t)\hat{B}_i(0)\rhoB\right],
\end{equation}
where
\begin{equation}
\hat{B}_i(t)=e^{i\HB t}\hat{B}_i e^{-i\HB t}.
\end{equation}
Without loss of generality, we can assume
\begin{equation}\label{eq:Bi_eq}
\TrB[\hat{B}_i\rhoB]=0.
\end{equation}
The Fourier transform of $\Phi_i(t)$ is denoted by $J_i(\omega)$:
\begin{equation}
J_i(\omega)=\int_{-\infty}^\infty \Phi(t)e^{-i\omega t}dt.
\end{equation}
In an infinitely large bath, which we consider in the following analysis, $J_i(\omega)$ is a continuous function of $\omega$ and $\Phi_i(t)$ vanishes in the long-time limit.

Following \citet{Nathan2020}, we introduce a jump correlator $g_i(t)$ as
\begin{equation}
g_i(t)=\int_{-\infty}^\infty \frac{d\omega}{2\pi}\sqrt{J_i(\omega)}e^{i\omega t},
\end{equation}
which is equivalent to
\begin{equation}\label{eq:Phi_g}
\Phi_i(t-t')=\int_{-\infty}^\infty ds\, g_i(t-s)g_i(s-t').
\end{equation}
We introduce two positive constants $\gamma$ and $\tauB$ satisfying
\begin{equation}\label{eq:ineq_g}
|g_i(t)|\leq \frac{\sqrt{\gamma}}{2\tauB}e^{-2|t|/\tauB}
\end{equation}
for arbitrary $i\in\Lambda$ and $t\in\mathbb{R}$.
From \cref{eq:ineq_g}, we can derive the following inequality for $\Phi_i(t)$:
\begin{equation}\label{eq:ineq_Phi}
|\Phi_i(t)|\leq\frac{\gamma}{\tauB}e^{-|t|/\tauB} \quad \forall i\in\Lambda.
\end{equation}
Physically, $\gamma$ and $\tauB$ denote the strength of dissipation and the bath correlation time, respectively.
It should be remarked that $\tauB$ should satisfy $\tauB\gtrsim \max_{i\in\Lambda}\beta_i$, which is a consequence of the Kubo-Martin-Schwinger condition~\citep{Mori2024_strong}: $J_i(\omega)=J_i(-\omega)e^{-\beta_i\omega}$.
The weak-coupling limit corresponds to the limit of $\gamma\to +0$ with $\tauB$ held fixed, whereas the singular-coupling limit corresponds to the limit of $\tauB\to +0$ with $\gamma$ held fixed.

Suppose a factorized initial state $\rhoT(0)=\rho(0)\otimes\rhoB$, where $\rhoT(t)$ is the density matrix of the total system and $\rho(t)$ is the reduced density matrix of the system of interest, which is defined as $\rho(t)=\TrB [\rhoT(t)]$.
We assume that $\rhoB$ is given by \cref{eq:rhoB}, i.e., each bath is in thermal equilibrium but its temperature can depend on site $i$.

The density matrix of the total system obeys the Liouville-von Neumann equation
\begin{equation}\label{eq:LvN}
\ddv{\rhoT(t)}{t}=-i[\hat{H}_T,\rho_T(t)].
\end{equation}
Formally, $\rhoT(t)$ is expressed as
\begin{equation}\label{eq:rhoT_t}
\rhoT(t)=\mathcal{U}(t)\rhoT(0),
\end{equation}
where $\mathcal{U}(t)(\cdot)=e^{i\HT t}(\cdot)e^{-i\HT t}$.
The exact reduced density matrix at time $t$ is thus given by $\rho(t)=\TrB[\mathcal{U}(t)\rho(0)\otimes\rhoB]$, which is in general non-Markovian.
By using various approximations (see \cref{sec:derivation} for technical details), all of which are expected to be valid as long as $\gamma\tauB\ll 1$, \citet{Nathan2020} derived the following ULE:
\begin{align}\label{eq:ULE}
\ddv{\rho}{t}
&=-i[\HS+\hat{\Delta},\rho]+\sum_{i=1}^N\left(\hat{L}_i\rho\hat{L}_i^\dagger-\frac{1}{2}\{\hat{L}_i^\dagger\hat{L}_i,\rho\}\right)
\nonumber \\
&\eqqcolon\mathcal{L}\rho, 
\end{align}
where $\mathcal{L}$ is the Liouvillian of the Lindblad form (it is also referred to as the Lindbladian).
Here,
\begin{equation}\label{eq:jump}
\hat{L}_i=\int_{-\infty}^\infty ds\, g_i(s)\hat{A}_i(-s)
\end{equation}
is the jump operator at site $i$ and $\hat{\Delta}=\sum_{i\in\Lambda}\hat{\Delta}_i$, where
\begin{equation}\label{eq:LS}
\hat{\Delta}_i=\frac{1}{2i}\int_{-\infty}^\infty ds\int_{-\infty}^\infty ds'\, \mathrm{sgn}(s-s')
g(s)g(-s')\hat{A}_i(s)\hat{A}_i(s'),
\end{equation}
is the modification of the system Hamiltonian as a result of the coupling to the heat bath, which is called the Lamb-shift Hamiltonian~\citep{Breuer_text}.
In both of \cref{eq:jump,eq:LS}, $\hat{A}_i(s)=e^{i\HS s}\hat{A}_ie^{-i\HS s}$.
In \cref{sec:derivation}, we briefly review the derivation of \cref{eq:ULE} following Ref.~\citep{Nathan2020}.

For later convenience, we decompose $\mathcal{L}$ into the unitary time evolution under $\HS$ and the effect of the coupling to the environment:
\begin{equation}
\mathcal{L}=\LS+\tilde{\mathcal{L}},
\end{equation}
where $\LS\coloneqq -i[\HS,\cdot]$. $\tilde{\mathcal{L}}$ is further decomposed into the Lamb-shift part $\mathcal{L}_\Delta$ and the dissipative part $\mathcal{D}$:
\begin{equation}
\left\{\begin{aligned}
&\tilde{\mathcal{L}}=\mathcal{L}_\Delta+\mathcal{D}, \\
&\mathcal{L}_\Delta\rho=-i[\hat{\Delta},\rho]=-i\sum_{i\in\Lambda}[\hat{\Delta}_i,\rho]\eqqcolon \sum_{i\in\Lambda}\mathcal{L}_{\Delta,i}, \\
&\mathcal{D}\rho=\sum_{i\in\Lambda}\left(\hat{L}_i\rho\hat{L}_i^\dagger-\frac{1}{2}\{\hat{L}_i^\dagger\hat{L}_i,\rho\}\right)\eqqcolon \sum_{i\in\Lambda}\mathcal{D}_i.
\end{aligned}\right.
\end{equation}

Let us introduce the adjoint $\mathcal{L}^\dagger$ of $\mathcal{L}$, which is expressed as
\begin{align}\label{eq:ULE_ad}
\mathcal{L}^\dagger\hat{O}&=i[\hat{H}_S+\hat{\Delta},\hat{O}]+\sum_{i=1}^N\left(\hat{L}_i^\dagger\hat{O}\hat{L}_i-\frac{1}{2}\{\hat{L}_i^\dagger\hat{L}_i,\hat{O}\}\right)
\nonumber \\
&=\LS^\dagger\hat{O}+\tilde{\mathcal{L}}^\dagger\hat{O}
\nonumber \\
&=\LS^\dagger\hat{O}+\mathcal{L}_\Delta^\dagger\hat{O}+\mathcal{D}^\dagger\hat{O}.
\end{align} 
The superoperator $\mathcal{L}^\dagger$ gives the time evolution of an operator in the Heisenberg picture: $\TrS[\hat{O}e^{\mathcal{L}t}\rho]=\TrS[(e^{\mathcal{L}^\dagger t}\hat{O})\rho]$.
It should be noted that $e^{\mathcal{L}^\dagger t}$ is a unital map: $\mathcal{L}^\dagger \hat{1}=0$, where $\hat{1}$ is the identity operator.

\subsection{Main result}
\label{sec:main}

We consider the time evolution of a local operator $\hat{O}_X$ that acts nontrivially to $X\subset\Lambda$, where $|X|$ is finite ($|X|$ denotes the number of elements of the set $X$).
Without loss of generality, we can put $\|\hat{O}_X\|=1$.

We compare the exact time evolution of the expectation value $\braket{\hat{O}_X(t)}\coloneqq\TrS[\hat{O}_X\rho(t)]$, where $\rho(t)=\TrB[\mathcal{U}(t)\rho(0)\otimes\rhoB]$ is the exact reduced density matrix at time $t$, with the approximate time evolution obeying the ULE: $\braket{\hat{O}_X(t)}_\mathrm{ULE}\coloneqq\TrS[\hat{O}_Xe^{\mathcal{L}t}\rho(0)]$.
We denote by $\epsilon(t)$ the difference between the two:
\begin{equation}\label{eq:epsilon}
\epsilon(t)\coloneqq \left|\braket{\hat{O}_X(t)}-\braket{\hat{O}_X(t)}_\mathrm{ULE}\right|,
\end{equation}
which is interpreted as the error of the ULE.

What we are going to show is that $\epsilon(t)$ is bounded from above by a constant $\bar{\epsilon}$, which depends on $\gamma$ and $\tauB$ but neither on time $t$ nor on the initial state $\rho(0)$, in the thermodynamic limit.
Moreover, $\bar{\epsilon}$ vanishes in the limit of $\gamma\to +0$ or $\tauB\to +0$.

By introducing dimensionless constants $\tilde{\gamma}=\gamma/v$ and $\tildetauB=v\tauB$, where $v$ denotes the Lieb-Robinson velocity (it will be defined in \cref{sec:LR}), it turns out that $\bar{\epsilon}=O(\tilde{\gamma}^{1/2}\tildetauB)$ for small $\gamma$ or small $\tauB$, which precisely means that $\bar{\epsilon}\sim \tilde{\gamma}^{1/2}$ for $\tilde{\gamma}\ll 1$ when $\tildetauB$ is held fixed and $\bar{\epsilon}\sim\tildetauB$ for $\tildetauB\ll 1$ when $\tilde{\gamma}$ is held fixed.
Thus, our main result is summarized as follows:
\begin{equation}
\lim_{N\to\infty}\epsilon(t)\leq\bar{\epsilon}=O(\tilde{\gamma}^{1/2}\tildetauB).
\end{equation}
For a sufficiently small value of $\gamma$ or $\tauB$, the ULE is justified for arbitrarily long times even in the thermodynamic limit.

In a non-rigorous but physically sound derivation of the ULE~\citep{Nathan2020} (see also Refs.~\citep{Mori2024_strong,Shiraishi2024}), it appears that the ULE is solely derived from the condition $\gamma\tauB\ll 1$.
However, the main result mentioned above implies that the condition of $\gamma\tauB\ll 1$ is not sufficient: a stronger condition $\tilde{\gamma}^{1/2}\tildetauB=v^{1/2}\gamma^{1/2}\tilde\tauB\ll 1$ is required for small $\tilde{\gamma}$.

\subsection{Theoretical methods}
\label{sec:method}

We briefly explain key theoretical methods: the Lieb-Robinson bound and the accelerated dissipation.

The Lieb-Robinson bound is a mathematical bound on the speed of propagation of information in spin systems.
It was originally proved for the unitary dynamics generated by a local Hamiltonian~\citep{Lieb1972,Hastings2004,Bravyi2006,Nachtergaele2006}, and later it was extended to Markovian open quantum systems~\citep{Nachtergaele2011,Poulin2010,Barthel2012}.
\Citet{Poulin2010} proved the Lieb-Robinson bound for open quantum systems described by a strictly local Lindbladian.
\Citet{Nachtergaele2011} extended it to exponentially decaying ones.

The ULE generator of \cref{eq:ULE} is, as it is, not written as a sum of exponentially decaying local terms, and hence it is not immediately obvious whether one can apply the result in Ref.~\citep{Nachtergaele2011} to the ULE.
In \cref{sec:LR}, we show that the Lieb-Robinson bound is valid for the ULE by showing that it can be further decomposed into exponentially decaying local terms.

The accelerated dissipation states that, under bulk dissipation, certain kinds of norms of a local operator decay at a rate increasing with time~\citep{Schuster2023,Shirai2024_accelerated,Mori2024_Liouvillian}.
Eventually, the decay rate becomes saturated at a value much larger than the initial value.
It occurs because of the combination of the operator spreading in the unitary time evolution~\citep{Nahum2018,Keyserlingk2018} and the presence of bulk dissipation.
Although the accelerated dissipation is considered to be a generic phenomenon, it has not been given as a rigorous theorem.

In this paper, we \emph{assume} the accelerated dissipation in the form of \cref{eq:accelerated_decay_no} or \cref{eq:accelerated_decay_diag,eq:accelerated_decay,eq:accelerated_decay_local}, depending on the situation.
We explain the statement of the accelerated dissipation and its intuitive picture in \cref{sec:time-dependent,sec:static}, and verify it numerically in \cref{sec:numerical}.
A large decay rate predicted by the accelerated dissipation is crucial for obtaining an error bound that is uniform in time.

\begin{figure}[t]
\centering
\includegraphics[width=0.9\linewidth]{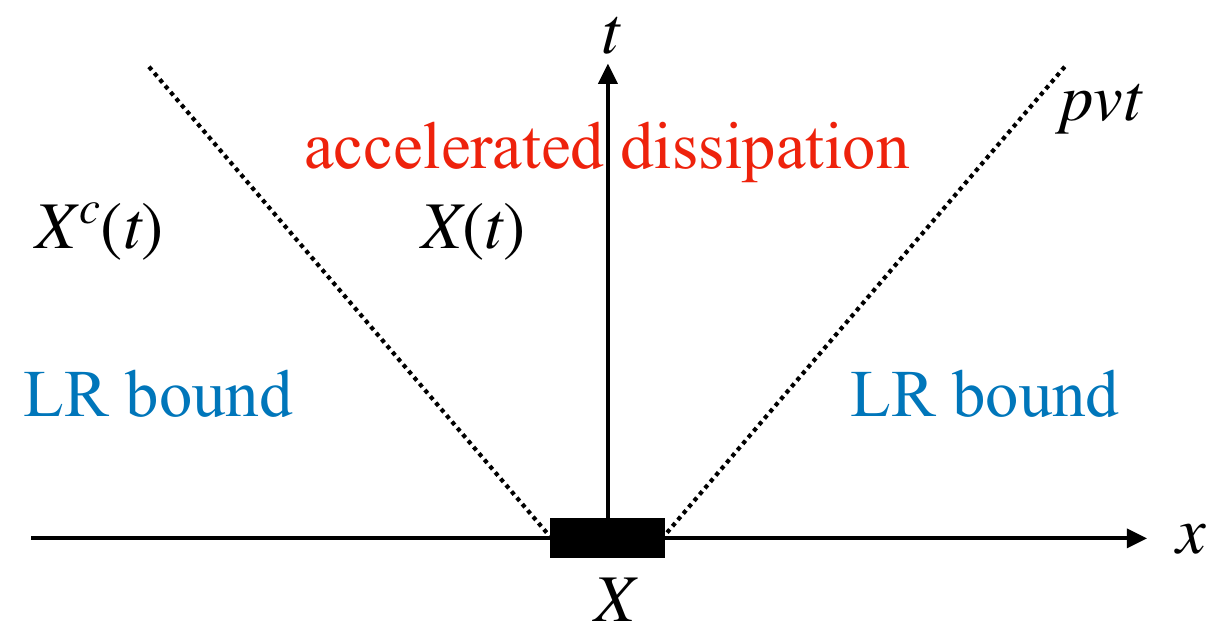}
\caption{Strategy to evaluate the error bound. Consider a local operator acting nontrivially onto $X$.
After time evolution, the operator grows. To evaluate the error bound, we use the Lieb-Robinson bound for the contribution from outside the light cone $X^c(t)$, whereas we use the accelerated dissipation for the contribution from inside the light cone $X(t)$.}
\label{fig:strategy}
\end{figure}

\Cref{fig:strategy} shows our strategy to evaluate the error bound.
As we will see later in \cref{sec:derivation}, the error from each approximation is expressed as a space-time summation.
We decompose the spacetime into inside and outside of the light cone $x=\pm pv|t|$, where $p>2$ is a constant and $v$ is the Lieb-Robinson velocity.
Outside the light cone, we use the Lieb-Robinson bound.
Inside the light cone, instead, we use the accelerated dissipation.

\section{Lieb-Robinson bound for the ULE}
\label{sec:LR}

Let us consider a time-dependent Lindbladian $\mathcal{L}(t)$, whose adjoint $\mathcal{L}^\dagger(t)$ is written as a sum of local terms:
\begin{equation}\label{eq:local_terms}
\mathcal{L}^\dagger(t)=\sum_{Z\subseteq\Lambda}\mathcal{L}_Z^\dagger(t),
\end{equation}
where $\mathcal{L}_Z^\dagger(t)$ is a superoperator with support $Z$, i.e. $\mathcal{L}_Z^\dagger(t)\hat{O}_X=0$ for any operator $\hat{O}_X$ whose support $X$ does not have intersection with $Z$: $X\cap Z=\emptyset$. 
For each $Z$, $\mathcal{L}_Z^\dagger(t)$ satisfies $\mathcal{L}_Z^\dagger(t)\hat{1}=0$, but it is not necessary to assume that $\mathcal{L}_Z^\dagger (t)$ is the adjoint of a Lindbladian in its own right.

We assume that the Lindbladian is exponentially decaying, which implies
\begin{equation}\label{eq:exp_decay}
\sum_{Z\supseteq\{i,j\}}\|\mathcal{L}_Z^\dagger(t)\|_\mathrm{cb}\leq De^{-\mu d(i,j)}
\end{equation}
for any $t$, where $\|\cdot\|_\mathrm{cb}$ denotes the completely bounded norm~\citep{Nachtergaele2011,Sweke2019}, and $D$ and $\mu$ are positive constants that are independent of the system size $N$.
Here, $d(i,j)$ is the distance between sites $i$ and $j$: $d(i,j)=|i-j|$ in the open boundary condition and $d(i,j)=\min\{|i-j|,N-|i-j|\}$ in the periodic boundary condition.

For the operator $\hat{O}_X$ whose support is $X$, let us define its dissipative time evolution in the Heisenberg picture as
\begin{equation}
\hat{O}_X^\mathrm{d}(t)=\bar{\mathcal{T}}\exp\left(\int_0^t ds\, \mathcal{L}^\dagger(s)\right)\hat{O}_X,
\end{equation}
where $\bar{\mathcal{T}}$ is the anti-time-ordering operation.
We here put the superscript ``d'' to distinguish the unitary time evolution $\hat{O}_X(t)=e^{i\HS t}\hat{O}e^{-i\HS t}$.


Under the conditions of \cref{eq:local_terms,eq:exp_decay}, the following Lieb-Robinson bound has been proved~\citep{Nachtergaele2011}:
\begin{align}\label{eq:LR}
\|[\hat{O}_X^\mathrm{d}(t),\hat{O}_Y]\|\leq C\|\hat{O}_X\|\|\hat{O}_Y\|\min\{|X|,|Y|\}
\nonumber \\
\times e^{-[d(X,Y)-vt]/\xi},
\end{align}
where $C$, $v$, $\xi$ are positive constants that depend on the lattice geometry and constants $D$ and $\mu$ in \cref{eq:exp_decay} but not on the system size $N$.
Here, $d(X,Y)=\min\{d(i,j): i\in X, j\in Y\}$ is the distance between the regions $X$ and $Y$.

It should be noted that the adjoint of the ULE Liouvillian, which is given by \cref{eq:ULE_ad} with \cref{eq:jump,eq:LS}, is \emph{not} in the form of \cref{eq:local_terms} satisfying \cref{eq:exp_decay}.
It is therefore not obvious whether one can apply the Lieb-Robinson bound for the ULE.
In the following, we show that the adjoint of the ULE Liouvillian can be brought into the form of \cref{eq:local_terms} satisfying \cref{eq:exp_decay}.
It means that the Lieb-Robinson bound holds for Markovian time evolution obeying the ULE.

In the expression $\mathcal{L}^\dagger=\LS^\dagger+\mathcal{L}_\Delta^\dagger+\mathcal{D}^\dagger$, $\LS^\dagger$ is expressed in the desired form because $\HS$ is a local Hamiltonian by assumption.
Therefore, the remaining task is to show that $\mathcal{L}_\Delta^\dagger$ and $\mathcal{D}^\dagger$ can be decomposed into exponentially decaying local terms.

The operators $\hat{\Delta}_i$ and $\hat{L}_i$ in the ULE are not strictly local because $\hat{A}_i(t)$ in \cref{eq:jump,eq:LS} become nonlocal for large $|t|$.
Considering the fact that the jump correlation function $g(t)$ vanishes for $|t|\gg\tauB$, it is expected that $\hat{A}_i(t)$ in \cref{eq:jump,eq:LS} can be approximated by a strictly local operator $\hat{A}_i^l(t)$. Here, we define $\hat{A}_i^l(t)$ as the restriction of $\hat{A}_i(t)$ to the region $R_{i,l}\coloneqq\{j\in\Lambda: d(i,j)\leq l\}$.
More precisely, we define
\begin{equation}
\hat{A}_i^l(t)=\frac{1}{\Tr_{R_{i,l}^c}[\hat{1}_{R_{i,l}^c}]}\Tr_{R_{i,l}^c}[\hat{A}_i(t)]\otimes\hat{1}_{R_{i,l}^c},
\end{equation}
where $\hat{1}_{R_{i,l}^c}$ is the identity operator for the region $R_{i,l}^c=\{j\in\Lambda: d(i,j)>l\}$.
The operator $\hat{A}_i^l(t)$ is regarded as a local operator approximating $\hat{A}_i(t)$.
By using the Lieb-Robinson bound for the unitary time evolution and recalling that $\hat{A}_i$ is a single-site operator with $\|\hat{A}_i\|=1$, we have
\begin{equation}\label{eq:LR_local}
\|\hat{A}_i(t)-\hat{A}_i^l(t)\|\leq \min\{2,Ce^{-(l-v|t|)/\xi}\}.
\end{equation}

Let us define $\hat{L}_i^l$ as the jump operator localized to the region $R_{i,l}$ by replacing $\hat{A}_i(-s)$ in \cref{eq:jump} by $\hat{A}_i^l(-s)$.
We also define $\hat{K}_i^l\coloneqq \hat{L}_i^l-\hat{L}_i^{l-1}$ for $l=0,1,\dots$ with $\hat{L}_i^{-1}\coloneqq 0$.
By using \cref{eq:LR_local,eq:ineq_g}, it is shown that the operator norm of $\hat{L}_i$ and $\hat{L}_i-\hat{L}_i^l$ are bounded from above as
\begin{equation}\label{eq:jump_norm}
\|\hat{L}_i^l\|\leq\|\hat{L}_i\|\leq\int_{-\infty}^\infty ds\,|g_i(s)|\leq\frac{\sqrt{\gamma}}{2},
\end{equation}
and
\begin{equation}\label{eq:jump_localize}
\|\hat{L}_i-\hat{L}_i^l\|\leq\frac{\sqrt{\gamma}}{2}(C+2)e^{-l/\eta},
\end{equation}
where
\begin{equation}
\eta=\max\{2\xi,v\tauB\}.
\end{equation}
This inequality leads to the following bound:
\begin{align}\label{eq:K_norm}
\|\hat{K}_i^l\|&\leq \|\hat{L}_i-\hat{L}_i^l\|+\|\hat{L}_i-\hat{L}_i^{l-1}\|
\nonumber \\
&\leq \sqrt{\gamma}(C+2)e^{-(l-1)/\eta}.
\end{align}
See \cref{sec:LR_ineq} for the detail on the derivation of \cref{eq:jump_localize}.

The dissipator $\mathcal{D}^\dagger=\sum_{i\in\Lambda}\mathcal{D}_i^\dagger$ is further decomposed into local terms as $\sum_{i\in\Lambda}\sum_l\mathcal{D}_{i,l}^\dagger$ with
\begin{align}\label{eq:D_il}
\mathcal{D}_{i,l}^\dagger\hat{O}=\hat{K}_i^{l\dagger}\hat{O}\hat{L}_i^l-\frac{1}{2}\{\hat{K}_i^{l\dagger}\hat{L}_i^l,\hat{O}\}+&\hat{L}_i^{l\dagger}\hat{O}\hat{K}_i^l-\frac{1}{2}\{\hat{L}_i^{l\dagger}\hat{K}_i^l,\hat{O}\}
\nonumber \\ 
-\hat{K}_i^{l\dagger}\hat{O}\hat{K}_i^l+\frac{1}{2}\{\hat{K}_i^{l\dagger}\hat{K}_i^l,\hat{O}\}.&
\end{align}
The completely bounded norm of $\mathcal{D}_{i,l}^\dagger$ is evaluated as
\begin{equation}
\|\mathcal{D}_{i,l}^\dagger\|_\mathrm{cb}\leq 4\|\hat{L}_i^l\|\|\hat{K}_i^l\|+2\|\hat{K}_i^l\|^2.
\end{equation}
By using \cref{eq:jump_norm,eq:K_norm}, we obtain
\begin{equation}\label{eq:norm_D_il}
\|\mathcal{D}_{i,l}^\dagger\|_\mathrm{cb}\leq 2\gamma(C+2)e^{-(l-1)/\eta}
+\gamma(C+2)^2e^{-2(l-1)/\eta},
\end{equation}
which shows that the dissipator of the ULE is exponentially decaying.

The Lamb-shift Hamiltonian $\hat{\Delta}$ is also decomposed into local terms as $\hat{\Delta}=\sum_{i\in\Lambda}\sum_l\hat{\Delta}_{i,l}$, where
\begin{align}\label{eq:Delta_il}
&\hat{\Delta}_{i,l}=\frac{1}{2i}\int_{-\infty}^\infty ds\int_{-\infty}^\infty ds'\, \mathrm{sgn}(s-s')g_i(s)g_i(-s')
\nonumber \\
&\times \left[(\hat{A}_i^l(s)-\hat{A}_i^{l-1}(s))\hat{A}_i^l(s')+\hat{A}_i^l(s)(\hat{A}_i^l(s')-\hat{A}_i^{l-1}(s'))\right].
\end{align}
Its operator norm is bounded as
\begin{align}\label{eq:norm_Delta_il}
&\|\hat{\Delta}_{i,l}\| \leq\frac{1}{2}\int_{-\infty}^\infty ds\int_{-\infty}^\infty ds' \, |g_i(s)||g_i(-s')|
\nonumber \\
&\qquad\times\left(\|\hat{A}_i^l(s)-\hat{A}_i^{l-1}(s)\|+\|\hat{A}_i^l(s')-\hat{A}_i^{l-1}(s')\|\right)
\nonumber \\
&\leq\frac{\gamma}{4\tauB}\int_{-\infty}^\infty ds\,e^{-2|s|/\tauB}\|\hat{A}_i^l(s)-\hat{A}_i^{l-1}(s)\|
\nonumber \\
&\leq \frac{\gamma}{2}(C+2)e^{-(l-1)/\eta}.
\end{align}
This bound implies that the Lamb-shift Hamiltonian has exponentially decaying interactions.

In conclusion, the adjoint of the generator of the ULE can be written as a sum of exponentially decaying local terms.
Therefore, we can apply the Lieb-Robinson bound of \cref{eq:LR} to the time evolution obeying the ULE.

\section{Accelerated dissipation}
\label{sec:acc}

\subsection{Systems with no local conserved quantity}
\label{sec:time-dependent}

Let us consider the dissipative time evolution of a local operator $\hat{O}$ in the Heisenberg picture: $\hat{O}^\mathrm{d}(t)=e^{\mathcal{L}^\dagger t}\hat{O}$.
We assume that $\mathcal{L}^\dagger$ is written in the form of \cref{eq:ULE_ad} with $\|\hat{L}_i\|\sim \sqrt{\gamma}$.
We also assume that each $\hat{L}_i$ is approximated by a local operator acting sites near $i$ (in the case of the ULE, \cref{eq:jump_localize} ensures this property).
Without loss of generality, we can assume that the expectation value of $\hat{O}$ in the steady state is zero, which implies $\lim_{t\to\infty}\hat{O}^\mathrm{d}(t)=0$.

First, suppose that the system of interest has no local conserved quantity when it is thermally isolated from the environment.
The system Hamiltonian must depend on time (if $\HS$ does not depend on time, the Hamiltonian $\HS$ itself is always a local conserved quantity).
Floquet systems, where the Hamiltonian periodically depends on time, and random quantum circuits are typical examples.
In such a case, the statement of the accelerated dissipation is given as follows:

\begin{statement}[systems with no conserved quantity]
\textit{Suppose that the time-dependent system Hamiltonian $\HS(t)$ has no local conserved quantity. Under the setup explained above, for any $\gamma^*>0$, there exist positive constants $a$ and $\zeta$ such that
\begin{equation}\label{eq:accelerated_decay_no}
\lim_{N\to\infty}\|\hat{O}^\mathrm{d}(t)\|\leq \zeta \|\hat{O}\|e^{-a\sqrt{\tilde{\gamma}}vt}
\end{equation}
for any $t>0$ and $\gamma\in(0,\gamma^*)$.}
\end{statement}

In the following, we explain an intuitive picture behind this statement.
Under bulk dissipation of strength $\gamma$, the decay rate of an operator $\hat{O}^\mathrm{d}(t)$ is roughly proportional to $\gamma \mathcal{S}(t)$, where $\mathcal{S}(t)$ is the average operator size of $\hat{O}^\mathrm{d}(t)$ (i.e. how many sites this operator acts on)~\citep{Schuster2023,Shirai2024_accelerated,Mori2024_Liouvillian}.
If the operator $\hat{O}$ has no overlap with all the powers of the Hamiltonian $\HS$, i.e. $\hat{O}$ has no diagonal elements in the energy basis, 
$\mathcal{S}(t)$ roughly obeys the following equation~\citep{Schuster2023,Shirai2024_accelerated}:
\begin{equation}\label{eq:size_EOM}
\frac{d\mathcal{S}(t)}{dt}\sim v-2\gamma\delta\mathcal{S}(t)^2,
\end{equation}
where $\delta\mathcal{S}(t)^2$ is the variance of the operator size.
The first term on the right-hand side implies that the operator size increases linearly $\mathcal{S}(t)\sim vt$, which is called operator spreading or operator growth~\citep{Keyserlingk2018,Nahum2018}.
The second term shows that bulk dissipation decreases the operator size at a rate proportional to the variance of the operator size.

When there is no local conserved quantity at $\gamma=0$, which can happen for a time-dependent Hamiltonian $\HS(t)$, the operator size distribution evolves diffusively $\delta\mathcal{S}(t)\sim\sqrt{t}$~\citep{Schuster2023}, and thus the operator size reaches the peak value $\mathcal{S}_\mathrm{peak}\sim\tilde{\gamma}^{-1}$, and then starts to decrease.
In this case, $\|\hat{O}^\mathrm{d}(t)\|\sim e^{-\gamma vt^2}$ up to $t\sim\gamma^{-1}$, and then the operator eventually undergoes an exponential decay $\|\hat{O}^\mathrm{d}(t)\|\sim e^{-gt}$ for $t\gg\tilde{\gamma}^{-1}$.
Here, $g$ is the spectral gap of the Liouvillian, which, interestingly, converges to a non-zero value in the weak dissipation limit after taking the thermodynamic limit: $\lim_{\gamma\to +0}\lim_{N\to\infty}g=\bar{g}>0$.
This finite decay rate is related to the intrinsic relaxation of the system, which is referred to as the quantum Ruelle-Pollicott resonance~\citep{Mori2024_Liouvillian,Prosen2002,Znidaric2024,Jacoby2025,Zhang2024,Yoshimura2025}.
It is thus expected that \cref{eq:accelerated_decay_no} holds for a time-dependent system Hamiltonian with no local conserved quantity.

The essential point is that a local operator decays in a timescale that is proportional to $\tilde{\gamma}^{-1/2}$ (see \cref{eq:accelerated_decay_no}), which is much shorter than a naively expected timescale $\tilde{\gamma}^{-1}$ for $\tilde{\gamma}\ll 1$. 
This is the meaning of the accelerated dissipation.

\subsection{Static system Hamiltonian}
\label{sec:static}

When there is a local conserved quantity at $\gamma=0$ (the Hamiltonian $\HS$ is always conserved in a static system), the presence of slow hydrodynamic modes alters the operator dissipation process.
Suppose a static Hamiltonian $\HS=\sum_{x\in\Lambda}\hat{h}_x$, where $\hat{h}_x$ is a local operator acting to sites around $x$.
For simplicity, we assume that there is no local conserved quantity apart from $\HS$.
We decompose the local operator $\hat{O}$ onto the diagonal part $\hat{O}_\mathrm{diag}$ and the off-diagonal part $\hat{O}_\mathrm{off}$ in the basis that diagonalizes $\HS$ (i.e. the energy basis).
After dissipative time evolution in the Heisenberg picture we have
\begin{equation}\label{eq:diag-off}
\hat{O}^\mathrm{d}(t)=\hat{O}_\mathrm{diag}(t)+\hat{O}_\mathrm{off}(t),
\end{equation}
where $\hat{O}_\mathrm{diag}(t)\coloneqq e^{\mathcal{L}^\dagger t}\hat{O}_\mathrm{diag}$ and $\hat{O}_\mathrm{off}(t)\coloneqq e^{\mathcal{L}^\dagger t}\hat{O}_\mathrm{off}$.
The diagonal part corresponds to the overlap with conserved quantities at $\gamma=0$.

As for the diagonal part, we expect that $\lim_{N\to\infty}\|[\hat{O}_\mathrm{diag}(t),\hat{A}(t')]\|=0$ for any local operator $\hat{A}$ and any times $t$ and $t'$, where $\hat{A}(t)=e^{i\HS t}\hat{A}e^{-i\HS t}$.
To understand it, suppose that the system of interest obeys the eigenstate thermalization hypothesis (ETH)~\citep{DAlessio2016_review,Mori2018_review}, which implies
\begin{equation}
\hat{O}_\mathrm{diag}=O_\mathrm{eq}(\HS/N)
\end{equation}
up to an exponentially small error in $N$, where $O_\mathrm{eq}(\cdot)$ is a smooth function.
Physically, $O_\mathrm{eq}(\varepsilon)$ corresponds to the equilibrium average of $\hat{O}$ at the energy density $\varepsilon$.
We then have $\|[\hat{O}_\mathrm{diag},\hat{A}(t')]\|=\|[O_\mathrm{eq}(\HS/N),\hat{A}]\|$, where we have used the fact that $\hat{O}_\mathrm{diag}$ commutes with $\HS$.
Performing the Taylor expansion of $O_\mathrm{eq}(\HS/N)$ as
\begin{equation}
O_\mathrm{eq}(\HS/N)=\sum_{n=0}^\infty\frac{O_\mathrm{eq}^{(n)}(0)}{n!}\left(\frac{\HS}{N}\right)^n,
\end{equation}
and using the inequality
\begin{equation}
\|[\HS^n,\hat{A}]\|\leq n\|\HS\|^{n-1}\|[\HS,\hat{A}]\|,
\end{equation}
we obtain
\begin{equation}\label{eq:comm_bound}
\|[O_\mathrm{eq}(\HS/N),\hat{A}]\|\lesssim \frac{\|[\HS,\hat{A}]\|}{N}.
\end{equation}
Here, we have assumed that the Taylor expansion of $dO_\mathrm{eq}(x)/dx$ around $x=0$ converges absolutely, and a non-essential constant factor is dropped in \cref{eq:comm_bound}.
When $\HS$ is a local Hamiltonian, $\|[\HS,\hat{A}]\|$ does not depend on the system size for any local operator $\hat{A}$, which implies $\|[\hat{O}_\mathrm{diag},\hat{A}(t')]\|\lesssim 1/N$, and hence the contribution from the conserved quantity (i.e. the diagonal part) is negligible in the thermodynamic limit.
It should be emphasized that the operator norm $\|\hat{O}_\mathrm{diag}\|$ does not vanish in the thermodynamic limit.
Taking the commutator with a local operator is crucial.
We expect that $\|[\hat{O}_\mathrm{diag}(t),\hat{A}(t')]\|$ is exponentially decreasing with $t$, and hence
\begin{equation}
\lim_{N\to\infty}\|[\hat{O}_\mathrm{diag}(t),\hat{A}(t')]\|=0.
\end{equation}
In this way, the contribution from the overlap with conserved quantities can be dropped in the thermodynamic limit when the commutator with a local operator is concerned.

Next, let us discuss how to decay the off-diagonal part $\hat{O}_\mathrm{off}(t)$.
The recent studies~\citep{Khemani2018,Rakovszky2018,Schuster2023} on the operator growth with conserved quantities revealed that a local operator in the Heisenberg picture at time $t$ is decomposed into \emph{the hydrodynamic components}, which are a linear combination of $\{\hat{h}_x\}_{x\in\Lambda}$, and \emph{the chaotic components}:
\begin{equation}\label{eq:hydro_chaos}
\hat{O}_\mathrm{off}(t)=\sum_{x\in\Lambda}\hat{O}_\mathrm{off}^{(x)}(t)=\sum_{x\in\Lambda}\left[c_x(t)\hat{h}_x+\hat{O}_x'(t)\right].
\end{equation}
The hydrodynamic components $\{c_x(t)\}$ spread diffusively and emit chaotic components, which spread ballistically.
$\hat{O}_x'(t)$ in \cref{eq:hydro_chaos} is the chaotic component emitted from the site $x$.
Under bulk dissipation of strength $\tilde{\gamma}$, the operator size of $\hat{O}_x'(t)$ is saturated at a certain value that is proportional to $\tilde{\gamma}^{-1/2}$~\citep{Schuster2023}.
We denote by $\ell_\gamma$ the saturated operator size of $\hat{O}_x'(t)$.
The decay rate is proportional to $\gamma\cdot\ell_\gamma\sim\tilde{\gamma}^{1/2}$.
Up to the saturation, the decay of the chaotic component is dominant, and hence we expect that the operator decays as
\begin{equation}
\sum_{x\in\Lambda}\|\hat{O}_\mathrm{off}^{(x)}(t)\|\leq\|\hat{O}\|\zeta e^{-a\tilde{\gamma}^{1/2}vt}
\end{equation}
with positive constants $\zeta$ and $a$.

Since the size of the operator $\hat{O}_\mathrm{off}^{(x)}(t)$ is at most $\ell_\gamma$, we have
\begin{equation}
[\hat{O}_\mathrm{off}^{(x)}(t),\hat{A}_i]\approx 0 \text{ when }d(i,x)>\ell_\gamma,
\end{equation}
where $\hat{A}_i$ is an operator acting to the site $i$.

The chaotic components decay faster than the hydrodynamic components, and hence, after saturation, the decay of $\|\hat{O}_\mathrm{off}^{(x)}(t)\|$ is dominated by that of the hydrodynamic components.
Since the size of the hydrodynamic component is $O(1)$, we expect that the decay rate is proportional to $\gamma$:
\begin{equation}\label{eq:chaos_long}
\sum_{x\in\Lambda}\|\hat{O}_\mathrm{off}^{(x)}(t)\|\leq \|\hat{O}\|\zeta'\tilde{\gamma}e^{-a'\gamma t}
\end{equation}
with positive constants $\zeta'$ and $a'$.
The amplitude of the asymptotic exponential decay of the hydrodynamic mode can depend on $\tilde{\gamma}$.
It is numerically found that it decreases with decreasing $\gamma$.
The prefactor $\tilde{\gamma}$ in \cref{eq:chaos_long} represents this dependence.

In summary, when $\HS$ is a static Hamiltonian (i.e. time-independent), the statement of the accelerated dissipation is given as follows:

\begin{statement}[static system Hamiltonian]
The time-evolved operator is decomposed into the diagonal and the off-diagonal parts as in \cref{eq:diag-off}.
As for the diagonal part, the commutator with a local operator vanishes in the thermodynamic limit:
\begin{equation}\label{eq:accelerated_decay_diag}
\lim_{N\to\infty}\|[\hat{O}_\mathrm{diag}(t),\hat{A}(t')]\|=0.
\end{equation}
As for the off-diagonal part, it is further decomposed into the sum of $\hat{O}_\mathrm{off}^{(x)}(t)$ as in \cref{eq:hydro_chaos}, where $\{\hat{O}_\mathrm{off}^{(x)}(t)\}$ satisfy the following properties:
(i) For any $\gamma^*>0$, there exist positive constants $a, a', \zeta, \zeta'$ such that
\begin{equation}\label{eq:accelerated_decay}
\lim_{N\to\infty}\sum_{x\in\Lambda}\|\hat{O}_\mathrm{off}^{(x)}(t)\|\leq\|\hat{O}\|\left(\zeta e^{-a\tilde{\gamma}^{1/2}vt}+\zeta'\tilde{\gamma}e^{-a'\gamma t}\right)
\end{equation}
for any $t>0$ and $\gamma\in(0,\gamma^*)$.
(ii) $\hat{O}_\mathrm{off}^{(x)}(t)$ is approximately given by an operator acting nontrivially onto the region $\{i\in\Lambda:d(i,x)<\ell_\gamma\}$ with $\ell_\gamma\sim\tilde{\gamma}^{-1/2}$, which implies
\begin{equation}\label{eq:accelerated_decay_local}
\|[\hat{O}_\mathrm{off}^{(x)}(t),\hat{A}_i]\|\approx 0 \,\, \text{when }d(i,x)\geq\ell_\gamma.
\end{equation}
\end{statement}

\subsection{Numerical verification}
\label{sec:numerical}


\begin{figure}[t]
\centering
\includegraphics[width=0.9\columnwidth]{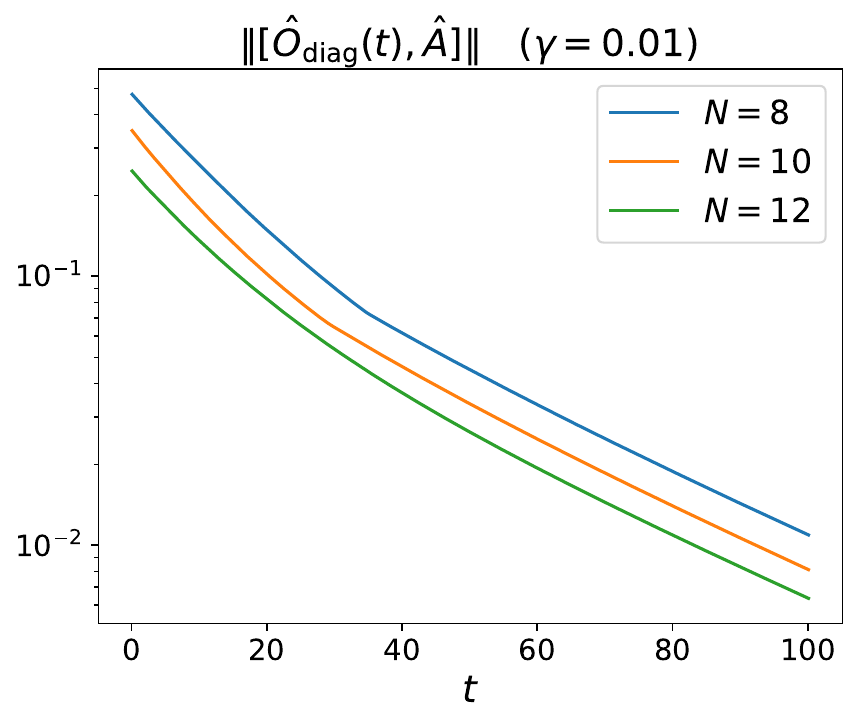}
\caption{Numerical test of \cref{eq:accelerated_decay_diag} for the bulk-dissipated Ising model in a tilted field. The time evolution of $\|[\hat{O}_\mathrm{diag}(t),\hat{A}]\|$ is plotted for system sizes $N=8, 10$, and 12. We choose $\hat{O}=\hat{\sigma}_1^x$ and $\hat{A}=\hat{\sigma}_1^z$. The dissipation strength is set as $\gamma=0.01$. We find that the quantity decreases as the system size increases, which is in accordance with \cref{eq:accelerated_decay_diag}.}
\label{fig:diag}
\end{figure}

To support the theoretical argument in \cref{sec:time-dependent,sec:static}, we numerically test the statement of the accelerated dissipation.
As it has already been mentioned, for a time-dependent system Hamiltonian with no local conserved quantity, recent studies on the quantum Ruelle-Pollicott resonance have shown that we have a finite decay rate of an operator in the limit of $\gamma\to 0$ after the thermodynamic limit, which is confirmed by numerical calculations~\citep{Mori2024_Liouvillian} as well as by analytic calculations for specific models~\citep{Yoshimura2024}.
It means that we already have convincing support for \cref{eq:accelerated_decay_no}, although there is no general proof.
In contrast, for a static Hamiltonian, we have no numerical or analytical results  that can support the validity of the statement of the accelerated dissipation.
Therefore we here focus on the numerical test of \cref{eq:accelerated_decay_diag,eq:accelerated_decay,eq:accelerated_decay_local} for a static system Hamiltonian.

Strictly speaking, these statements are formulated in the thermodynamic limit, but numerical calculations are limited to relatively small system sizes owing to an exponentially large dimensionality of the operator space.
We therefore have to be content with the numerical test up to the largest system size ($N=12$) we can access.


Let us consider the Ising model in a tilted field:
\begin{equation}
\HS=-\sum_{i=1}^N\left(\hat{\sigma}_i^z\hat{\sigma}_{i+1}^z+h_x\hat{\sigma}_i^x+h_z\hat{\sigma}_i^z\right),
\end{equation}
where $\hat{\sigma}_i^\alpha$ ($\alpha=x,y,z$) denote the Pauli matrices at site $i$.
For both cases, we consider the periodic boundary condition and set $h_x=0.9045$ and $h_z=0.8090$ following \citep{Kim2014}, for which the eigenstate thermalization hypothesis is numerically verified.
Bulk dissipation is modeled by the two jump operators $\hat{L}_i^+=\sqrt{\gamma}\hat{\sigma}_i^+$ and $\hat{L}_i^-=\sqrt{\gamma}\hat{\sigma}_i^-$ at each site $i$.
The strength of the nearest-neighbor interaction is put unity, which fixes the unit of the local energy scale of the system.
Therefore, we do not have to distinguish $\gamma$ and $\tilde{\gamma}$ in numerical calculations.

First, we start with the numerical test of \cref{eq:accelerated_decay_diag}.
We choose $\hat{O}=\hat{\sigma}_1^x$ and $\hat{A}=\hat{\sigma}_1^z$, and compute $\|[\hat{O}_\mathrm{diag}(t),\hat{A}]\|$ for system sizes $N=8, 10$, and $12$.
The numerical results are plotted in \cref{fig:diag}.
We see that the norm decreases as the system size increases, which is in accordance with \cref{eq:accelerated_decay_diag}.
In \cref{sec:static}, it is argued that \cref{eq:accelerated_decay_diag} can be understood as a consequence of the ETH.
So far, we have many numerical examples of evidence on the validity of the ETH for various nonintegrable models~\citep{Mori2018_review}, which implies that \cref{eq:accelerated_decay_diag} has indirectly been validated by previous works.

\begin{figure*}[t]
\centering
\includegraphics[width=0.9\linewidth]{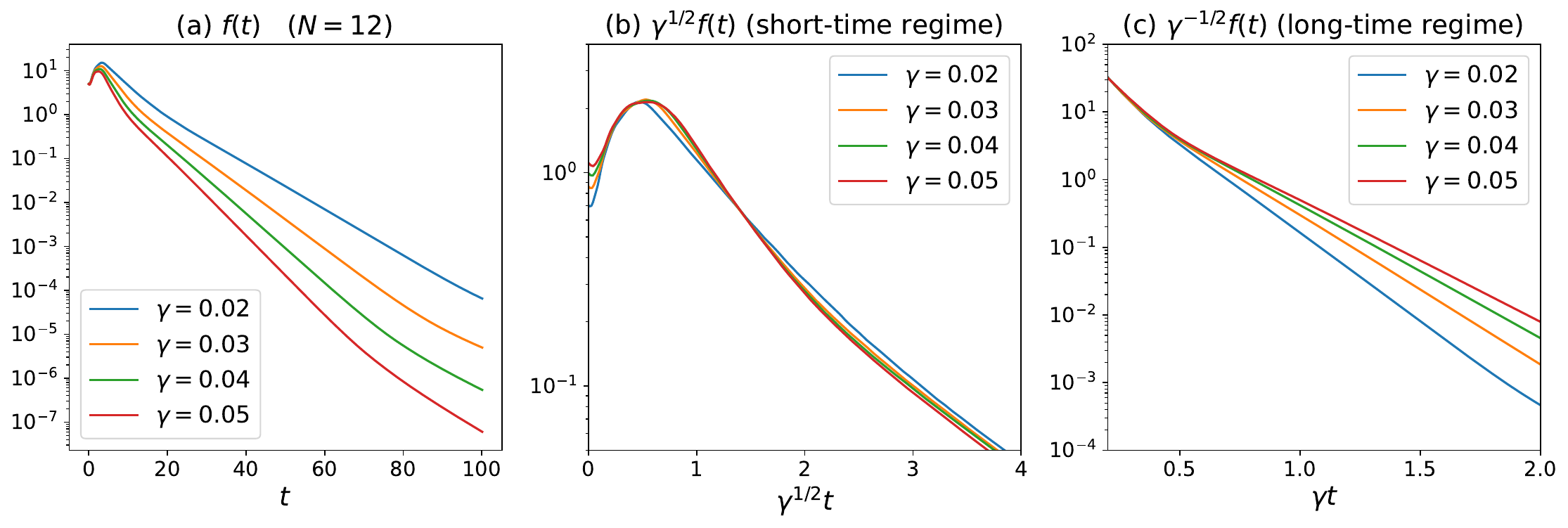}
\caption{Plot of $f(t)$ for $\hat{O}=\hat{\sigma}_1^x$ and $\hat{A}_i=\hat{\sigma}_i^z$. In all figures, the system size is set as $N=12$. (a) The time evolution of $f(t)$. $f(t)$ shows a peak at small $t$ and then exhibits exponential decay. (b) The short-time regime. $\gamma^{1/2}f(t)$ is plotted against $\gamma^{1/2}t$. It is found that the peak value of $f(t)$ is $O(\gamma^{-1/2})$, and the decay rate of $f(t)$ after the peak is proportional to $\gamma^{1/2}$, which is in accordance with the first term of \cref{eq:f}. (c) The long-time regime. $\gamma^{-1/2}f(t)$ is plotted against $\gamma t$. It is found that $f(t)$ exponentially decays, and moreover, at a fixed value of $\gamma t$, $\gamma^{-1/2}f(t)$ decreases as decreasing $\gamma$. This observation is in accordance with the second term of \cref{eq:f}.}
\label{fig:f}
\end{figure*}

Next, we consider \cref{eq:accelerated_decay,eq:accelerated_decay_local}.
It is difficult to directly test them, we instead consider the following quantity:
\begin{equation}
f(t)\coloneqq \lim_{N\to\infty}\sum_{i\in\Lambda}\|[\hat{O}_\mathrm{off}(t),\hat{A}_i]\|,
\end{equation}
where $\hat{A}_i$ is a local operator at site $i$ that satisfies $\|\hat{A}_i\|\leq 1$.
By using \cref{eq:accelerated_decay,eq:accelerated_decay_local}, $f(t)$ is evaluated as follows:
\begin{align}\label{eq:f}
f(t)&\leq\lim_{N\to\infty}\sum_{i\in\Lambda}\sum_{x\in\Lambda}\|[\hat{O}_\mathrm{off}^{(x)}(t),\hat{A}_i]\|
\nonumber \\
&\approx\lim_{N\to\infty}\sum_{x\in\Lambda}\sum_{i: d(i,x)\leq\ell_\gamma}\|[\hat{O}_\mathrm{off}^{(x)}(t),\hat{A}_i]\|
\nonumber \\
&\lesssim \ell_\gamma\lim_{N\to\infty}\sum_{x\in\Lambda}\|[\hat{O}_\mathrm{off}^{(x)}(t)\|
\nonumber \\
&\lesssim \|\hat{O}\|\left(\zeta\tilde{\gamma}^{-1/2}e^{-a\tilde{\gamma}^{1/2}vt}+\zeta'\tilde{\gamma}^{1/2}e^{-a'\gamma t}\right).
\end{align}
We use \cref{eq:accelerated_decay_local} at the second line and \cref{eq:accelerated_decay} at the last line.

We compute $f(t)$ for $\hat{O}=\hat{\sigma}_1^x$ and $\hat{A}_i=\hat{\sigma}_i^z$, and check whether the inequality (\ref{eq:f}) holds for small $\gamma$.
The upper bound in \cref{eq:f} consists of two parts: the contribution from the short-time regime $f(t)\lesssim\tilde{\gamma}^{-1/2}e^{-a\tilde{\gamma}^{1/2}vt}$ and that from the long-time regime $f(t)\lesssim\tilde{\gamma}^{1/2}e^{-a'\gamma t}$.
We numerically test them separately.

In \cref{fig:f} (a), we plot $f(t)$ for the system size $N=12$ for various values of $\gamma$.
For small $t$, $f(t)$ grows, which corresponds to the increase of the operator size of the chaotic components $\{\hat{O}_x'(t)\}$.
After the peak, it decreases exponentially.
In \cref{fig:f}(b), we plot $\gamma^{1/2}f(t)$ as a function of $\gamma^{1/2}t$, which corresponds to the short-time regime.
We can see that the peak value of $\gamma^{1/2}f(t)$ does not increase as decreasing $\gamma$, and $\gamma^{1/2}f(t)$ shows an exponential decay $e^{-a\gamma^{1/2}t}$ after the peak (i.e. when plotted on the semi-log axes as a function of $\gamma^{1/2}t$, the slope is almost independent of $\gamma$). 
These numerical observations imply that the inequality $f(t)\leq\zeta\tilde{\gamma}^{-1/2}e^{-a\tilde{\gamma}^{1/2}vt}$ seems to hold when properly choosing constants $\zeta$ and $a$.
In \cref{fig:f}(c), we plot $\gamma^{-1/2}f(t)$ as a function of $\gamma t$, which corresponds to the long-time regime.
We can see that for any fixed time, $\gamma^{-1/2}f(t)$ decreases as decreasing $\gamma$, which verifies the inequality $f(t)\leq\zeta'\tilde{\gamma}^{1/2}e^{-a'\gamma t}$.

Theoretically, the statement of the accelerated dissipation, i.e. \cref{eq:accelerated_decay_no} or \cref{eq:accelerated_decay_diag,eq:accelerated_decay,eq:accelerated_decay_local}, is still a conjecture, but numerical results show that it is plausible.

%

\section{Error bounds}
\label{sec:error}

In this section, we derive the main result summarized in \cref{sec:summary}.
In \cref{sec:derivation}, we begin with a brief summary of the derivation of the ULE, where every source of errors is mathematically expressed.
By using those mathematical expressions, in \cref{sec:bound}, we illustrate how to use the theoretical methods described in \cref{sec:method} to evaluate the error bound.
Full details are given in \cref{sec:detail}.

\subsection{Preliminary: derivation of the ULE}
\label{sec:derivation}

In this section, we review the derivation of \cref{eq:ULE} following Ref.~\citep{Nathan2020}.
Through the derivation, we give the explicit expression of the error term for each approximation, which is used to evaluate the error bounds on the dynamics of a local quantity in \cref{sec:error}.

Let us start with the Liouville-von Neumann equation for the total system, i.e., \cref{eq:LvN}, and its formal solution given by \cref{eq:rhoT_t} with a factorized initial state $\rhoT(0)=\rho(0)\otimes\rhoB$.
In the interaction picture, $\rhoT^\mathrm{I}(t)=e^{i(\HS+\HB)t}\rhoT(t)e^{-i(\HS+\HB)t}$, which is formally written as
\begin{equation}\label{eq:rhoTI_t}
\rhoT^\mathrm{I}(t)=\mathcal{U}_\mathrm{I}(t,0)\rhoT(0), \quad \mathcal{U}_\mathrm{I}(t,s)=\mathcal{T}e^{\int_s^tds'\, \LI(s')},
\end{equation}
where $\mathcal{T}$ is the time-ordering operator and $\LI(t)=-i[\HI(t),\cdot]$.
Here, $\HI(t)$ is the interaction Hamiltonian in the interaction picture, which is given by
\begin{equation}
\HI(t)=\sum_{i\in\Lambda}\hat{A}_i(t)\otimes\hat{B}_i(t),
\end{equation}
where $\hat{A}_i(t)=e^{i\HS t}\hat{A}_ie^{-i\HS t}$ and $\hat{B}_i(t)=e^{i\HB t}\hat{B}_ie^{-i\HB t}$.

In the interaction picture, we have
\begin{equation}
\frac{d\rhoT^\mathrm{I}(t)}{dt}=\LI(t)\rhoT^\mathrm{I}(t)=\LI(t)\mathcal{U}_\mathrm{I}(t,0)\rho(0)\otimes\rhoB.
\end{equation}
The reduced density matrix in the interaction picture is given by $\rho^\mathrm{I}(t)=\TrB[\rhoT^\mathrm{I}(t)]$.
By using Wick's theorem for free bosons, we have
\begin{align}\label{eq:drhoIdt_Wick}
&\frac{d\rho^\mathrm{I}(t)}{dt}=\TrB[\LI(t)\mathcal{U}^\mathrm{I}(t,0)\rho(0)\otimes\rhoB]
\nonumber \\
&=-\sum_{i\in\Lambda}\sum_{\nu=\pm 1}\nu\int_0^t ds\, \Phi_i(\nu(t-s))
\nonumber \\
&\times [\hat{A}_i(t),\TrB[\mathcal{U}_\mathrm{I}(t,s)\mathcal{A}_i^\nu(s)\mathcal{U}_\mathrm{I}(s,0)\rho(0)\otimes\rhoB]],
\end{align}
where $\mathcal{A}_i^{+1}(s)(\cdot)=\hat{A}_i(s)(\cdot)$ and $\mathcal{A}_i^{-1}(s)(\cdot)=(\cdot)\hat{A}_i(s)$. 
That is, we multiply $\hat{A}_i(s)$ from left when $\nu=1$ and from right when $\nu=-1$.

Here, $\mathcal{U}_\mathrm{I}(t,s)$ is rewritten as
\begin{equation}
\mathcal{U}_\mathrm{I}(t,s)=1+\int_s^tds'\, \LI(s')\mathcal{U}_\mathrm{I}(s',s).
\end{equation}
By substituting it into \cref{eq:drhoIdt_Wick}, we have
\begin{align}\label{eq:QME_B}
\frac{d\rho^\mathrm{I}(t)}{dt}=-\sum_{i\in\Lambda}\sum_{\nu=\pm 1}\nu\int_0^t ds\, \Phi_i(\nu(t-s))
\nonumber \\
\times [\hat{A}_i(t),\mathcal{A}_i^\nu(s)\rho^\mathrm{I}(s)]
+\hat{\xi}_\mathrm{B}^\mathrm{I}(t),
\end{align}
where
\begin{align}\label{eq:xi_B}
\hat{\xi}_\mathrm{B}^\mathrm{I}(t)=i\sum_{i,j\in\Lambda}\sum_{\nu=\pm 1}\nu\int_0^tds\,\Phi_i(\nu(t-s))\int_s^tds'
\nonumber \\
[\hat{A}_i(t),[\hat{A}_j(s'),\Psi_{ij}^\nu(s',s)]],
\end{align}
where
\begin{equation}
\Psi_{ij}^\nu(s',s)=\TrB[\hat{B}_j(s')\mathcal{U}_\mathrm{I}(s',s)\mathcal{A}_i^\nu(s)\mathcal{U}_\mathrm{I}(s,0)\rho(0)\otimes\rhoB].
\end{equation}
Ignoring $\hat{\xi}_\mathrm{B}^\mathrm{I}(t)$ corresponds to the Born approximation.
Therefore, $\hat{\xi}_\mathrm{B}^\mathrm{I}(t)$ expresses the error induced by the Born approximation in the interaction picture.

Next, we perform the Markov approximation: in \cref{eq:QME_B}, replace $\rho^\mathrm{I}(s)$ by $\rho^\mathrm{I}(t)$ and $\int_0^tds$ by $\int_{-\infty}^tds$.
As a result, we obtain
\begin{align}\label{eq:QME_BM}
\frac{d\rho^\mathrm{I}(t)}{dt}=-\sum_{i\in\Lambda}\sum_{\nu=\pm 1}\nu&\int_0^\infty ds\, \Phi_i(\nu s)[\hat{A}_i(t),\mathcal{A}_i^\nu(t-s)\rho^\mathrm{I}(t)]
\nonumber \\
&+\hat{\xi}_\mathrm{B}^\mathrm{I}(t)+\hat{\xi}_\mathrm{M}^\mathrm{I}(t),
\end{align}
where the error induced by the Markov approximation is given by
\begin{align}\label{eq:xi_M}
\hat{\xi}_\mathrm{M}^\mathrm{I}(t)&=\sum_{i\in\Lambda}\sum_{\nu=\pm 1}\nu\int_0^t ds\, \Phi_i(\nu s)
\nonumber \\
&\times [\hat{A}_i(t),\mathcal{A}_i^\nu(t-s)(\rho^\mathrm{I}(t)-\rho^\mathrm{I}(t-s))]
\nonumber \\
&+\sum_{i\in\Lambda}\sum_{\nu=\pm 1}\nu\int_t^\infty ds\,\Phi_i(\nu s)[\hat{A}_i(t),\mathcal{A}_i^\nu(t-s)\rho^\mathrm{I}(t)].
\end{align}

By introducing an operator
\begin{equation}
\hat{R}_i(t)=\int_0^\infty ds\, \Phi_i(s)\hat{A}_i(t-s),
\end{equation}
and using $\Phi_i(-t)=\Phi_i(t)^*$, \cref{eq:QME_BM} is expressed as
\begin{equation}\label{eq:Redfield}
\frac{d\rho^\mathrm{I}}{dt}=-\sum_{i\in\Lambda}[\hat{A}_i(t),\hat{R}_i(t)\rho^\mathrm{I}-\rho^\mathrm{I}\hat{R}_i^\dagger(t)]+\hat{\xi}_\mathrm{BM}^\mathrm{I}(t),
\end{equation}
where $\hat{\xi}_\mathrm{BM}^\mathrm{I}(t)=\hat{\xi}_\mathrm{B}^\mathrm{I}(t)+\hat{\xi}_\mathrm{M}^\mathrm{I}(t)$.
\Cref{eq:Redfield} is known as the Redfield equation~\citep{Breuer_text} when we ignore $\hat{\xi}_\mathrm{BM}^\mathrm{I}(t)$.

In a textbook derivation of the Lindblad equation~\citep{Breuer_text}, the secular approximation is used, but it is in general not valid in many-body systems, as is mentioned in Introduction.
\Citet{Nathan2020,Nathan2024} introduce the memory-dressing transformation to derive the ULE, which is explained below.

Let us first rewrite \cref{eq:Redfield} by using the jump correlation function, see \cref{eq:Phi_g}.
By using a superoperator $\mathcal{F}(t,s,s')$, which is defined as
\begin{align}\label{eq:F}
\mathcal{F}(t,s,s')\hat{O}&=\sum_{i\in\Lambda}\theta(t-s')
\nonumber \\
&\times\left\{g_i(t-s)g_i(s-s')[\hat{A}_i(t),\hat{O}\hat{A}_i(s')]\right.
\nonumber \\
&\left.+g_i^*(t-s)g_i^*(s-s')[\hat{A}_i(s')\hat{O},\hat{A}_i(t)]\right\},
\end{align}
\cref{eq:Redfield} is expressed as follows:
\begin{equation}\label{eq:Redfield_F}
\frac{d\rho^\mathrm{I}}{dt}=\int_{-\infty}^\infty ds\int_{-\infty}^\infty ds'\,\mathcal{F}(t,s,s')\rho^\mathrm{I}(t)+\hat{\xi}_\mathrm{BM}^\mathrm{I}(t).
\end{equation}

A key observation by Nathan and Rudner is that if we replace $\mathcal{F}(t,s,s')$ by $\mathcal{F}(s,t,s')$, the above equation becomes of the Lindblad form, which is nothing but the ULE, except for the error term.
To make a connection between the Redfield equation and the ULE, let us introduce the memory-dressing transformation $\rho^\mathrm{I}(t)\to\tilde{\rho}^\mathrm{I}(t)$, which is defined as
\begin{equation}\label{eq:modified}
\tilde{\rho}^\mathrm{I}(t)=[1+\mathcal{M}(t)]\rho^\mathrm{I}(t),
\end{equation}
where
\begin{align}\label{eq:M}
\mathcal{M}(t)=\int_t^\infty ds_1&\int_{-\infty}^t ds_2\int_{-\infty}^\infty ds_3
\nonumber \\
&[\mathcal{F}(s_1,s_2,s_3)-\mathcal{F}(s_2,s_1,s_3)].
\end{align}
The transformed density matrix $\tilde{\rho}^\mathrm{I}(t)$ is referred to as the modified density matrix~\citep{Nathan2020}.
By differentiating \cref{eq:modified} with respect to $t$ and using \cref{eq:Redfield_F,eq:M}, we obtain
\begin{equation}\label{eq:EOM_modified}
\frac{d\tilde{\rho}^\mathrm{I}}{dt}=\int_{-\infty}^\infty ds\int_{-\infty}^\infty ds'\,\mathcal{F}(s,t,s')\rho^\mathrm{I}(t)+\hat{\xi}_\mathrm{BM}^\mathrm{I}(t)+\hat{\xi}_1^\mathrm{I}(t),
\end{equation}
where
\begin{equation}
\hat{\xi}_1^\mathrm{I}(t)=\mathcal{M}(t)\pd_t\rho^\mathrm{I}(t).
\end{equation}

Straightforward calculations show that the first term of the right hand side of \cref{eq:EOM_modified} is nothing but that of the ULE, i.e. \cref{eq:ULE}, in the interaction picture.
The ULE Liouvillian in the interaction picture is given by 
\begin{equation}
\tilde{\mathcal{L}}^\mathrm{I}(t)=e^{-\LS t}\mathcal{L}e^{\LS t}-\LS=e^{-\LS t}\tilde{\mathcal{L}}e^{\LS t},
\end{equation} 
where recall that we write $\mathcal{L}=\LS+\tilde{\mathcal{L}}$.
We have
\begin{equation}
\tilde{\mathcal{L}}^\mathrm{I}(t)=\int_{-\infty}^\infty ds\int_{-\infty}^\infty ds'\,\mathcal{F}(s,t,s').
\end{equation}
\Cref{eq:EOM_modified} is thus written as
\begin{equation}\label{eq:modified_ULE}
\frac{d\tilde{\rho}^\mathrm{I}}{dt}=\tilde{\mathcal{L}}^\mathrm{I}(t)\tilde{\rho}^\mathrm{I}(t)+\hat{\xi}_\mathrm{BM}^\mathrm{I}(t)+\hat{\xi}_1^\mathrm{I}(t)+\hat{\xi}_2^\mathrm{I}(t),
\end{equation}
where
\begin{equation}
\hat{\xi}_2^\mathrm{I}(t)=\tilde{\mathcal{L}}^\mathrm{I}(t)[\rho^\mathrm{I}(t)-\tilde{\rho}^\mathrm{I}(t)].
\end{equation}
The modified density matrix obeys the ULE when we ignore 
\begin{equation}
\hat{\xi}_\mathrm{all}^\mathrm{I}(t)\coloneqq \hat{\xi}_\mathrm{BM}^\mathrm{I}(t)+\hat{\xi}_1^\mathrm{I}(t)+\hat{\xi}_2^\mathrm{I}(t),
\end{equation}
which is interpreted as the total correction to the ULE for the modified density matrix.

If the modified density matrix $\tilde{\rho}^\mathrm{I}(t)$ is close to the exact density matrix $\rho^\mathrm{I}(t)$ and $\hat{\xi}_\mathrm{all}$ is small enough, $\rho^\mathrm{I}(t)$ approximately obeys the ULE: $d\rho(t)/dt\approx\mathcal{L}\rho(t)$ by going back to the Schr\"odinger picture.

In Ref.~\citep{Nathan2020}, the following upper bounds on the trace norm of $\hat{\xi}_\mathrm{all}^\mathrm{I}(t)$ and that of $\rho^\mathrm{I}(t)-\tilde{\rho}^\mathrm{I}(t)$ are obtained~\footnote{Actually, \citet{Nathan2020} considered the spectral norm, but we can repeat the same argument by using the trace norm without any change.}:
\begin{equation}\label{eq:NR_xi}
 \|\hat{\xi}_\mathrm{all}^\mathrm{I}(t)\|_1\lesssim (N\gamma)^2\tauB
\end{equation}
and
\begin{equation}\label{eq:NR_rho}
\|\rho^\mathrm{I}(t)-\tilde{\rho}^\mathrm{I}(t)\|_1\lesssim N\gamma\tauB,
\end{equation}
respectively, where $\|\cdot\|_1$ denotes the trace norm.
Since $\|\mathcal{L}^\mathrm{I}(t)\rho^\mathrm{I}(t)\|_1\lesssim \gamma N$, \cref{eq:NR_xi} implies that the correction term is actually small when $N\gamma\tauB\ll 1$.
Under the same condition, \cref{eq:NR_rho} shows that we can ignore the difference between the exact density matrix and the modified one.

However, there are two problems in the above evaluation as is already pointed out in Introduction.
Firstly, if we are interested in many-body systems under bulk dissipation, the condition $N\gamma\tauB\ll 1$ is not satisfied in the thermodynamic limit ($N\to\infty$).
Secondly, \cref{eq:NR_xi} gives an error bound at the level of the equations of motion, but errors might be accumulated with time.
It is important to exclude this possibility to ensure that the ULE can be used to predict long-time (i.e. $t\gtrsim 1/\gamma$) evolution of the system.

In the next section, we give an error bound on $\epsilon(t)$ that is given by \cref{eq:epsilon}.
It turns out that $\epsilon(t)$ keeps small during the time evolution in the thermodynamic limit when $\gamma$ or $\tauB$ is small enough.

\subsection{Decomposition of the total error}
\label{sec:bound}

Our purpose is to evaluate the upper bound on $\epsilon(t)=|\braket{\hat{O}_X(t)}-\braket{\hat{O}_X(t)}_\mathrm{ULE}|$, which is the difference between the exact expectation value of an operator $\hat{O}_X$ at time $t$ and the approximate one evaluated by the ULE.
We always consider the error in the thermodynamic limit, and hence we omit to put $\lim_{N\to\infty}$ in each equation hereafter.

We first decompose the total error $\epsilon(t)$ into three parts by using the triangle inequality:
\begin{align}\label{eq:error_decomp}
\epsilon(t)&\leq\|\TrS[\hat{O}_X(\rho(t)-\tilde{\rho}(t))]\|
\nonumber \\
&+\|\TrS[\hat{O}_X(\tilde{\rho}(t)-e^{\mathcal{L}t}\tilde{\rho}(0))]\|
\nonumber \\
&+\|\TrS[\hat{O}_Xe^{\mathcal{L}t}(\rho(0)-\tilde{\rho}(0))]\|
\end{align}
First term on the right-hand side represents the error from the difference between the exact density matrix $\rho(t)$ and the modified one in the Schr\"odinger picture, $\tilde{\rho}(t)=e^{-i\HS t}\tilde{\rho}^\mathrm{I}(t)e^{i\HS t}$.
The second term represents the error from the correction to the ULE for the modified density matrix, see \cref{eq:modified_ULE}.
The last term represents the error from the difference between the initial state of the exact density matrix and that of the modified one.

As for the second term of \cref{eq:error_decomp}, by using the Schr\"odinger picture of \cref{eq:modified_ULE}, i.e.
\begin{equation}
\frac{d\tilde{\rho}(t)}{dt}=\mathcal{L}\tilde{\rho}(t)+\hat{\xi}_\mathrm{all}(t),
\end{equation}
where $\hat{\xi}_\mathrm{all}(t)=e^{-i\HS t}\hat{\xi}_\mathrm{all}^\mathrm{I}(t)e^{i\HS t}$, we obtain
\begin{equation}
\tilde{\rho}(t)-e^{\mathcal{L}t}\tilde{\rho}(0)=\int_0^t dt'\, e^{\mathcal{L}t'}\hat{\xi}_\mathrm{all}(t-t').
\end{equation}
By further decomposing $\hat{\xi}_\mathrm{all}(t)$ into $\hat{\xi}_\mathrm{B}(t)+\hat{\xi}_\mathrm{M}(t)+\hat{\xi}_1(t)+\hat{\xi}_2(t)$, we have
\begin{align}
&\|\TrS[\hat{O}_X(\tilde{\rho}(t)-e^{\mathcal{L}t}\tilde{\rho}(0))]\|
\nonumber \\
&\leq \int_0^tdt'\,\left|\TrS[\hat{O}_Xe^{\mathcal{L}t'}\hat{\xi}_\mathrm{all}(t-t')\right|
\nonumber \\
&\leq\sum_{\alpha=\mathrm{B},\mathrm{M},1,2}\int_0^t dt'\, \left|\TrS[\hat{O}_X^\mathrm{d}(t')\hat{\xi}_\alpha(t-t')]\right|,
\end{align}
where recall that $\hat{O}_X^\mathrm{d}(t)=e^{\mathcal{L}^\dagger t}\hat{O}_X$.
In this way, we can decompose the total error into the error from each approximation: $\epsilon(t)\leq \epsilon_\mathrm{B}(t)+\epsilon_\mathrm{M}(t)+\epsilon_\mathrm{md}(t)$, where
\begin{align}
&\text{(Born approximation)}\nonumber \\
&\epsilon_\mathrm{B}(t)=\int_0^t dt'\, \left|\TrS[\hat{O}_X^\mathrm{d}(t')\hat{\xi}_\mathrm{B}(t-t')]\right| \\
&\text{(Markov approximation)}\nonumber \\
&\epsilon_\mathrm{M}(t)=\int_0^t dt'\, \left|\TrS[\hat{O}_X^\mathrm{d}(t')\hat{\xi}_\mathrm{M}(t-t')]\right| \\
&\text{(Memory-dressing transformation)}\nonumber \\
&\epsilon_\mathrm{md}(t)=\int_0^t dt'\, \sum_{\alpha=1,2}\left|\TrS[\hat{O}_X^\mathrm{d}(t')\hat{\xi}_\alpha(t-t')]\right|
\nonumber \\
&+\left|\TrS[\hat{O}_X(\rho(t)-\tilde{\rho}(t))]\right|+\left|\TrS[\hat{O}_X^\mathrm{d}(t)(\rho(0)-\tilde{\rho}(0))]\right|
\label{eq:epsilon_mod}
\end{align}

Now we shall evaluate an upper bound for each error term.
Instead of explaining full details here, we illustrate how to use the theoretical methods described in \cref{sec:method} (i.e. the Lieb-Robinson bound and the accelerated dissipation) to evaluate the error bound for the Born approximation $\epsilon_\mathrm{B}(t)$.
Full details are given in \cref{sec:detail}.

\subsection{Born approximation}

Our aim here is to evaluate the upper bound on $\epsilon_\mathrm{B}(t)$.
By using \cref{eq:xi_B} and the cyclic property of the trace, and changing the integration variables as $s\to t-t'-s$ and $s'\to t-t'-s'$, we obtain
\begin{align}
&\epsilon_\mathrm{B}(t)\leq\sum_{i,j\in\Lambda}\sum_{\nu=\pm 1}\int_0^t dt'\int_0^{t-t'}ds\,\int_0^sds'\,|\Phi_i(\nu s)|
\nonumber \\
&\times \|[[\hat{O}_X^\mathrm{d}(t'),\hat{A}_i],\hat{A}_j(-s')]\|\cdot \|\Psi_{ij}^\nu(t-t'-s',t-t'-s)\|_1.
\end{align}
By using Wick's theorem, \cref{eq:ineq_Phi}, and $\|\hat{A}_i\|=1$, we have
\begin{align}
&\|\Psi_{ij}^\nu(t-t'-s',t-t'-s)\|_1\leq 
\nonumber \\
&\sum_{\nu'=\pm 1}\int_{s'}^{t-t'}ds''\, |\Phi_j(\nu'(s'-s''))|
\nonumber \\
&\leq\frac{2\gamma}{\tauB}\int_{s'}^\infty ds''\, e^{-|s''-s'|/\tauB}=2\gamma.
\end{align}
By using \cref{eq:ineq_Phi} again, $\epsilon_\mathrm{B}(t)$ is bounded as
\begin{align}
\epsilon_\mathrm{B}(t)\leq \sum_{i,j\in\Lambda}\frac{4\gamma^2}{\tauB}\int_0^t dt'\int_0^{t-t'}ds\,\int_0^s ds'\, e^{-s/\tauB}
\nonumber \\
\times\|[[\hat{O}_X^\mathrm{d}(t'),\hat{A}_i],\hat{A}_j(-s')]\|.
\end{align}
Since the right hand side is an increasing function of $t$, we obtain
\begin{align}\label{eq:epsilon_B_upper}
\epsilon_\mathrm{B}(t)&\leq \sum_{i,j\in\Lambda}\frac{4\gamma^2}{\tauB}\int_0^\infty dt'\int_0^\infty ds\,\int_0^s ds'\, e^{-s/\tauB}
\nonumber \\
&\qquad\qquad\times\|[[\hat{O}_X^\mathrm{d}(t'),\hat{A}_i],\hat{A}_j(-s')]\|
\nonumber \\
&\eqqcolon \bar{\epsilon}_\mathrm{B}.
\end{align}
It should be noted that this upper bound does not depend on $t$.
Therefore, if it is shown that $\bar{\epsilon}_\mathrm{B}$ is sufficiently small, the Born approximation is justified for arbitrarily long times.

In the following analysis, we change the integration variable $t'$ to $t$ for notational simplicity.
There is no confusion about it because $\bar{\epsilon}_\mathrm{B}$ no longer depend on $t$.

We introduce a region $X(t)$ defined as
\begin{equation}
X(t)=\{i\in\Lambda: d(X,i)< pvt\},
\end{equation}
where $v$ is the Lieb-Robinson velocity and $p$ is an arbitrary constant satisfying $p>2$ (e.g. we can put $p=3$).
In the following analysis, the summation over $i$ and $j$ in \cref{eq:epsilon_B_upper} is decomposed into the three contributions: (i) $i,j\in X(t)$, (ii) $i\in X^c(t)$ and $d(X,i)\geq d(X,j)$, and (iii) $j\in X^c(t)$ and $d(X,j)>d(X,i)$.
Correspondingly, $\bar{\epsilon}_\mathrm{B}=\bar{\epsilon}_\mathrm{B}^\mathrm{(i)}+\bar{\epsilon}_\mathrm{B}^\mathrm{(ii)}+\bar{\epsilon}_\mathrm{B}^\mathrm{(iii)}$.
Below, we evaluate each contribution.
In the following analysis, it turns out that the case (i) gives the dominant contribution, and we have $\bar{\epsilon}_\mathrm{B}\lesssim\tilde{\gamma}^{1/2}\tilde{\tau}_\mathrm{B}$.

\subsubsection*{(i) $i,j\in X(t)$}

First, we consider a time-dependent system Hamiltonian $\HS(t)$ with no local conserved quantity.
In this case we can use \cref{eq:accelerated_decay_no} to evaluate the error bound.
From \cref{eq:epsilon_B_upper} with the restriction $i,j\in X(t)$, we have
\begin{align}
\bar{\epsilon}_\mathrm{B}^\mathrm{(i)}&\leq 16\gamma^2\tauB\int_0^\infty dt \sum_{i,j\in X(t)}\|\hat{O}_X^\mathrm{d}(t)\|
\nonumber \\
&\leq 16\gamma^2\tauB\int_0^\infty dt\, (|X|+2pvt)^2\|\hat{O}_X^\mathrm{d}(t)\|,
\end{align}
where we have used $\|[\hat{O}_X^\mathrm{d}(t),\hat{A}_i],\hat{A}_j(-s')]\|\leq 4\|\hat{O}_X^\mathrm{d}(t)\|$ and $|X(t)|\leq |X|+2pvt$.
By using \cref{eq:accelerated_decay_no}, we obtain
\begin{equation}
\bar{\epsilon}_\mathrm{B}^\mathrm{(i)}\leq16\tildetauB\left(\frac{8p^2}{a^3}\tilde{\gamma}^{1/2}+\frac{4|X|p}{a^2}\tilde{\gamma}+\frac{|X|^2}{a}\tilde{\gamma}^{3/2}\right),
\end{equation}
which is $O(\tilde{\gamma}^{1/2}\tildetauB)$ for small $\tilde{\gamma}$ or small $\tildetauB$.

Next, we consider a static system Hamiltonian $\HS$ in which \cref{eq:accelerated_decay_diag,eq:accelerated_decay,eq:accelerated_decay_local} hold.
We decompose $\hat{O}_X^\mathrm{d}(t)$ into the diagonal and the off-diagonal parts: $\hat{O}_X^\mathrm{d}(t)=\hat{O}_\mathrm{diag}(t)+\hat{O}_\mathrm{off}(t)$.
From \cref{eq:accelerated_decay_diag}, the contribution from $\hat{O}_\mathrm{diag}(t)$ can be ignored in the thermodynamic limit:
\begin{align}
&\lim_{N\to\infty}\sum_{i,j\in X(t)}\|[[\hat{O}_\mathrm{diag}(t),\hat{A}_i],\hat{A}_j(-s')]\|
\nonumber \\
&\leq 2|X(t)|^2\max_{i\in X(t)}\lim_{N\to\infty}\|[\hat{O}_\mathrm{diag}(t),\hat{A}_i]\|=0.
\end{align}
By using it and \cref{eq:hydro_chaos}, we have
\begin{equation}
\|[[\hat{O}_X^\mathrm{d}(t),\hat{A}_i],\hat{A}_j(-s')]\|\leq\sum_{x\in X(t)}\|[[\hat{O}_\mathrm{off}^{(x)}(t),\hat{A}_i],\hat{A}_j(-s')]\|.
\end{equation}
Here, recall that $\hat{O}_\mathrm{off}^{(x)}(t)$ is approximately a local operator acting nontrivially onto $R_x\coloneqq \{i\in\Lambda: d(i,x)< \ell_\gamma\}$, see \cref{eq:accelerated_decay_local}.
It means that $\sum_{i\in X(t)}$ can be replaced by $\sum_{i\in R_x}$:
\begin{align}
&\sum_{i,j,x\in X(t)}\|[[\hat{O}_\mathrm{off}^{(x)}(t),\hat{A}_i],\hat{A}_j(-s')]\|
\nonumber \\
&\approx\sum_{x,j\in X(t)}\sum_{i\in R_x}\|[[\hat{O}_\mathrm{off}^{(x)}(t),\hat{A}_i],\hat{A}_j(-s')]\|.
\end{align}
Since $[\hat{O}_\mathrm{off}^{(x)}(t),\hat{A}_i]$ is an operator acting nontrivially onto $R_x$, by using the Lieb-Robinson bound we have
\begin{align}\label{eq:B_piece1}
&\sum_{x,j\in X(t)}\sum_{i\in R_x}\|[[\hat{O}_\mathrm{off}^{(x)}(t),\hat{A}_i],\hat{A}_j(-s')]\|
\nonumber \\
&\leq \sum_{x,j\in X(t)}\sum_{i\in R_x}\|[\hat{O}_\mathrm{off}^{(x)}(t),\hat{A}_i]\|\max\{2,e^{-[d(R_x,j)-vs']/\xi}\}
\nonumber \\
&\leq4\ell_\gamma\left(\zeta e^{-a\tilde{\gamma}^{1/2}vt}+\zeta'\tilde{\gamma}e^{-a'\gamma t}\right)
\nonumber \\
&\qquad\times\sum_{j\in X(t)}\max\{2,e^{-[d(R_x,j)-vs']/\xi}\}.
\end{align}
We further evaluate $\sum_{j\in X(t)}\max\{2,e^{-[d(R_x,j)-vs']/\xi}\}$ as
\begin{align}\label{eq:B_piece2}
&\sum_{j\in X(t)}\max\{2,e^{-[d(R_x,j)-vs']/\xi}\}
\nonumber \\
&\leq\sum_{j:d(R_x,j)< vs'}2+\sum_{j:d(R_x,j)\geq vs'}e^{-[d(R_x,j)-vs']/\xi}
\nonumber \\
&\approx 4(vs'+\ell_\gamma)+2\int_{vs'}^\infty dr\, e^{-(r-vs')/\xi}
\nonumber \\
&=4(vs'+\ell_\gamma)+2\xi.
\end{align}
By substituting \cref{eq:B_piece1,eq:B_piece2} into \cref{eq:epsilon_B_upper} (with the restriction $i,j\in X(t)$), we have
\begin{align}
\bar{\epsilon}_\mathrm{B}^{(i)}&\leq \frac{32\ell_\gamma v^2}{\tildetauB}\left(\frac{\zeta}{a}\tilde{\gamma}^{3/2}+\frac{\zeta'}{a'}\tilde{\gamma}^2\right)
\nonumber \\
&\qquad\times\int_0^\infty ds\,e^{-s/\tauB}[vs^2+(2\ell_\gamma+\xi)s]
\nonumber \\
&=32\ell_\gamma\left(\frac{\zeta}{a}\tilde{\gamma}^{3/2}+\frac{\zeta'}{a'}\tilde{\gamma}^2\right)
[2\tildetauB^2+(2\ell_\gamma+\xi)\tildetauB].
\end{align}
By considering $\ell_\gamma\sim\tilde{\gamma}^{-1/2}$, we find $\bar{\epsilon}_\mathrm{B}^{(i)}\leq O(\tilde{\gamma}^{1/2}\tildetauB)$ for small $\tilde{\gamma}$ or $\tildetauB$.

It should be noted that if we used a milder assumption $\|\hat{O}_X^\mathrm{d}(t)\|\leq e^{-O(\gamma t)}$, we would obtain $\bar{\epsilon}_\mathrm{B}^\mathrm{(i)}\sim\tilde{\tau}_\mathrm{B}/\tilde{\gamma}$, which does not tend to zero in the weak-coupling limit $\tilde{\gamma}\to +0$.
To obtain the error bound that vanishes in the weak-coupling limit, the assumption of the accelerated decay, i.e. \cref{eq:accelerated_decay_no} or \cref{eq:accelerated_decay_diag,eq:accelerated_decay,eq:accelerated_decay_local}, is required.

\subsubsection*{(ii) $i\in X^c(t)$ and $d(X,i)\geq d(X,j)$}
By using 
\begin{align}
\|[[\hat{O}_X^\mathrm{d}(t),\hat{A}_i],\hat{A}_j(-s')]\|\leq 2\|[\hat{O}_X^\mathrm{d}(t),\hat{A}_i]\|
\nonumber \\
\leq 2Ce^{-[d(X,i)-vt]/\xi},
\end{align}
where the Lieb-Robinson bound (\cref{eq:LR}) is used in the last inequality, we have
\begin{align}\label{eq:epsilon_B_ii_upper}
\bar{\epsilon}_\mathrm{B}^\mathrm{(ii)}\leq 8C\gamma^2\tauB\int_0^\infty dt\,\sum_{i\in X^c(t)}\sum_{j: d(X,j)\leq d(X,i)}e^{-[d(X,i)-vt]/\xi}.
\end{align}
By putting $d(X,i)=r$, we have $\sum_{i\in X^c(t)}\approx 2\int_{pvt}^\infty dr$ and $\sum_{j:d(X,j)\leq d(X,i)}1=|X|+2r$.
We thus have
\begin{equation}
\bar{\epsilon}_\mathrm{B}^\mathrm{(ii)}\leq 16C\gamma^2\tauB\int_0^\infty dt\int_{pvt}^\infty dr(|X|+2r)
e^{-(r-vt)/\xi}.
\end{equation}
After carrying out the integrations, we obtain
\begin{equation}
\bar{\epsilon}_\mathrm{B}^\mathrm{(ii)}\leq 16C\frac{2(2p-1)\xi^3+(p-1)\xi^2|X|}{(p-1)^2}\tilde{\gamma}^2\tilde{\tau}_\mathrm{B},
\end{equation}
which is of order $\tilde{\gamma}^2\tilde{\tau}_\mathrm{B}$ and not dominant compared with $\bar{\epsilon}_\mathrm{B}^\mathrm{(i)}$ for $\tilde{\gamma}\ll 1$.

\subsubsection*{(iii) $j\in X^c(t)$ and $d(X,j)>d(X,i)$}

We introduce the region $X_j$ defined as
\begin{equation}\label{eq:Xj}
X_j=\left\{i\in\Lambda: d(X,i)\leq \frac{p-1}{p}d(X,j)\right\}.
\end{equation}
We then introduce the operator $\hat{O}_{X,j}^\mathrm{d}(t)$ as the operator obtained by localizing $\hat{O}_X^\mathrm{d)}(t)$ to the region $X_j$:
\begin{equation}\label{eq:O_Xj}
\hat{O}_{X,j}^\mathrm{d}(t)=\frac{\Tr_{X_j^c}[\hat{O}_X^\mathrm{d}(t)]\otimes\hat{1}_{X_j^c}}{\Tr_{X_j^c}[\hat{1}_{X_j^c}]},
\end{equation}
where $X_j^c$ is the complement of $X_j$.
Let us decompose $\|[[\hat{O}_X^\mathrm{d}(t),\hat{A}_i],\hat{A}_j(-s')]\|$ in $\bar{\epsilon}_\mathrm{B}^\mathrm{(iii)}$ as follows:
\begin{align}\label{eq:Born_iii}
\|[[\hat{O}_X^\mathrm{d}(t),\hat{A}_i],\hat{A}_j(-s')]\|
\leq \|[[\hat{O}_{X,j}^\mathrm{d}(t),\hat{A}_i],\hat{A}_j(-s')]\|
\nonumber \\
+\|[[\hat{O}_X^\mathrm{d}(t)-\hat{O}_{X,j}^\mathrm{d}(t),\hat{A}_i],\hat{A}_j(-s')]\|.
\end{align}
By using the Lieb-Robinson bound, we have
\begin{align}\label{eq:Born_iii_1}
&\|[[\hat{O}_{X,j}^\mathrm{d}(t),\hat{A}_i],\hat{A}_j(-s')]\|
\nonumber \\
&\leq \|[\hat{O}_{X,j}^\mathrm{d}(t),\hat{A}_i]\|\min\{2, Ce^{-[d(X_j,j)-vs']/\xi}\}
\nonumber \\
&\leq 2I[i\in X_j]\min\{2, Ce^{-[d(X_j,j)-vs']/\xi}\}
\nonumber \\
&\leq 2I[i\in X_j]\min\{2, Ce^{-\left[\frac{1}{p}d(X,j)-vs'\right]/\xi}\}
\end{align}
where $I[\cdot]$ is the indicator function, and
\begin{align}\label{eq:Born_iii_2}
&\|[[\hat{O}_X^\mathrm{d}(t)-\hat{O}_{X,j}^\mathrm{d}(t),\hat{A}_i],\hat{A}_j(-s')]\|
\leq 4\|\hat{O}_X^\mathrm{d}(t)-\hat{O}_{X,j}^\mathrm{d}(t)\|
\nonumber \\
&\qquad\leq 4C|X|e^{-\left[\frac{p-1}{p}d(X,j)-vt\right]/\xi}.
\end{align}
By substituting \cref{eq:Born_iii_1,eq:Born_iii_2} into \cref{eq:Born_iii}, we obtain
\begin{align}
&\bar{\epsilon}_\mathrm{B}^\mathrm{(iii)}
\leq \frac{4\gamma^2}{\tauB}\int_0^\infty dt\int_0^\infty ds\int_0^s ds'\,e^{-s/\tauB}\sum_{j\in X^c(t)}\sum_{i:d(X,i)<d(X,j)}
\nonumber \\
&\times\left[2I[i\in X_j]\min\left\{2,Ce^{-\left[\frac{1}{p}d(X,j)-vs'\right]/\xi}\right\}\right.
\nonumber \\
&\qquad\qquad\left.+4C|X|e^{-\left[\frac{p-1}{p}d(X,j)-vt\right]/\xi}\right].
\end{align}
As for $\min\{2,Ce^{-[\frac{1}{p}d(X,j)-vs']/\xi}\}$, we choose 2 when $pvt<d(X,j)<pvs'$, and choose $Ce^{-[\frac{1}{p}d(X,j)-vs']/\xi]}$ when $d(X,j)\geq pvs'$.
As a result, we have 
\begin{equation}
\bar{\epsilon}_\mathrm{B}^\mathrm{(iii)}\leq B_1+B_2+B_3,
\end{equation}
where
\begin{align}
B_1=\frac{16\gamma^2}{\tauB}\int_0^\infty dt\int_t^\infty ds\int_t^s ds'\,e^{-s/\tauB}\times
\nonumber \\
\sum_{j:pvt<d(X,j)<pvs'}\sum_{i\in X_j}1,
\end{align}
\begin{align}
B_2=\frac{8C\gamma^2}{\tauB}\int_0^\infty dt\int_0^\infty ds\int_0^sds'\, e^{-s/\tauB}\times
\nonumber \\
\sum_{j:d(X,j)>pv\max\{t,s'\}}\sum_{i\in X_j}e^{-\left[\frac{d(X,j)}{p}-vs'\right]/\xi},
\end{align}
and
\begin{align}
B_3=\frac{16C|X|\gamma^2}{\tauB}\int_0^\infty dt\int_0^\infty ds\int_0^sds'\, e^{-s/\tauB}\times
\nonumber \\
\sum_{j:d(X,j)>pvt}\sum_{i:d(X,i)<d(X,j)}e^{-\left[\frac{p-1}{p}d(X,j)-vt\right]/\xi}.
\end{align}

By introducing $r=d(X,j)$ and $r'=d(X,i)$, and replace the sum over $i$ and $j$ by integrations over $r$ and $r'$, we can explicitly evaluate $B_1$, $B_2$, and $B_3$.
We omit tedious calculations, and just present the result:
\begin{equation}
\begin{aligned}
&B_1\approx 32\tilde{\gamma}^2\tilde{\tau}_\mathrm{B}^3[p|X|+4p(p-1)\tilde{\tau}_\mathrm{B}],
\\
&B_2\approx 16C\tilde{\gamma}^2\tilde{\tau}_\mathrm{B}[p|X|\xi(\xi+\tilde{\tau}_\mathrm{B})+4p(p-1)\xi(\xi^2+\xi\tilde{\tau}_\mathrm{B}+\tilde{\tau}_\mathrm{B}^2)],
\\
&B_3\approx \frac{32C|X|\xi^2p}{(p-1)(p-2)}\tilde{\gamma}^2\tilde{\tau}_\mathrm{B}\left[|X|+\frac{2p(2p-3)}{(p-1)(p-2)}\xi\right].
\end{aligned}
\end{equation}
By collecting them, $\bar{\epsilon}_\mathrm{B}^\mathrm{(iii)}=O(\tilde{\gamma}^2\tilde{\tau}_\mathrm{B})$ for small $\tilde{\gamma}$ or $\tilde{\tau}_\mathrm{B}$, which is not dominant compared with $\bar{\epsilon}_\mathrm{B}^\mathrm{(i)}$.

In this way, the error bound for the Born approximation has been derived.
We decompose the set $\Lambda$ of lattice sites into $X(t)$, which corresponds to the inside of the light cone, and $X^c(t)$, which corresponds to the outside of the light cone.
Inside the light cone, we use the accelerated dissipation, and consequently obtain the error that is bounded by $O(\tilde{\gamma}^{1/2}\tildetauB)$.
Outside the light cone, we use the Lieb-Robinson bound, and obtain the error that is bounded by $O(\tilde{\gamma}^2\tildetauB)$.
The main source of error comes from the inside of the light cone.
Our error bound, $\epsilon(t)\lesssim\tilde{\gamma}^{1/2}\tildetauB$ is larger than expected: In a heuristic derivation we only assume $\gamma\tauB\ll 1$~\citep{Nathan2020,Mori2024_strong,Shiraishi2024}, and hence we expect $\epsilon(t)\sim\gamma\tauB$.
The reason why the error is enhanced is because of the operator growth: An operator with larger size is more fragile to bulk dissipation.

The evaluation of error bounds for the other approximations is done in essentially the same manner.

\section{Conclusion}
\label{sec:conclusion}

The ULE describes dissipative time evolution of an open quantum many-body system.
However, in its microscopic derivation, several approximations are introduced, and hence it has not been obvious whether the ULE gives accurate predictions, especially for the long-time behavior of macroscopically large quantum systems.
In this paper, we evaluate errors on the time evolution of local quantities.
For this purpose, we show that the Lieb-Robinson bound holds in an open system obeying the ULE, and introduce an assumption of accelerated dissipation, which is physically motivated by recent works on operator scrambling.
The Lieb-Robinson bound and the accelerated dissipation are key tools to evaluate error bounds.

We find that the error is bounded from above by a quantity of order $\tilde{\gamma}^{1/2}\tildetauB$, where $\tilde{\gamma}$ is the dimensionless dissipation strength and $\tildetauB$ is the dimensionless correlation time of the environment.
It is therefore concluded that either the weak-coupling limit $\tilde{\gamma}\to +0$ or the singular-coupling limit $\tildetauB\to +0$ justifies the use of the ULE for arbitrarily long times in the thermodynamic limit.

Dissipation usually destroys quantum coherence, and hence it is crucial to understand the effect of dissipation and develop techniques for reducing environmental effects in quantum technologies~\citep{Cai2023}.
Meanwhile, recent experimental progress~\citep{Bloch2012,Barreiro2011,Barontini2013,Tomita2017} introduces a new perspective: Engineered dissipation can control and manipulate quantum states~\citep{Verstraete2009,Harrington2022}.
The interplay between dissipation and interactions gives rise to rich nonequilibrium dynamics~\citep{Cai2013,Bouganne2020} and new kinds of quantum phases~\citep{Diehl2010,Klinder2015,Kessler2021}.
It is important to develop new theoretical tools and lay the foundation for them to tame dissipation in quantum many-body systems.
This work takes a step forward in this direction.

Finally, we conclude with open problems.
The accelerated dissipation is used as a key assumption in this work: it is yet to be formulated as a mathematical theorem.
It is an important future problem to refine the notion of the accelerated dissipation as a fundamental tool to describe open quantum many-body systems.
It is also a future work to justify the Lindblad equation in boundary dissipated quantum many-body systems.
The setup of boundary dissipation is relevant for quantum transport in a nonequilibrium steady state~\citep{Prosen2008,Prosen2011}.
In such a system, we typically observe delayed relaxation in the sense that the relaxation time is much larger than the inverse of the spectral gap of the Liouvillian~\citep{Znidaric2015,Mori2020_resolving}.
The delayed relaxation is considered to originate from the presence of bulk hydrodynamic modes, which do not decay until they reach the boundary.
It is then nontrivial how to control errors stemming from the inside of the light cone in \cref{fig:strategy}.
We hope to address these problems in the future.

\begin{acknowledgments}
This work was supported by JSPS KAKENHI Grant No. JP21H05185 and by JST, PRESTO Grant No. JPMJPR2259.
\end{acknowledgments}

\appendix

\section{Derivation of \cref{eq:jump_localize}}
\label{sec:LR_ineq}

The definition of the jump operator in the ULE is given in \cref{eq:jump}.
By using it, we have
\begin{equation}
\|\hat{L}_i-\hat{L}_i^l\|\leq\int_{-\infty}^\infty ds\, |g_i(s)|\|\hat{A}_i(-s)-\hat{A}_i^l(-s)\|.
\end{equation}
By using \cref{eq:ineq_g,eq:LR_local}, it is bounded as
\begin{align}
\|\hat{L}_i-\hat{L}_i^l\|&\leq\int_{-\infty}^\infty ds\, \frac{\sqrt{\gamma}}{2\tauB}e^{-2|s|/\tauB}
\min\{2,Ce^{-(l-v|s|)/\xi}\}
\nonumber \\
&=\frac{\sqrt{\gamma}}{\tauB}\int_0^\infty ds\, e^{-2s/\tauB}\min\{2,Ce^{-(l-vs)/\xi}\}.
\end{align}
Here, we decompose the integration range as $\int_0^\infty=\int_0^{l/2v}ds+\int_{l/2v}^\infty ds$, where we use $\min\{2,Ce^{-(l-vs)/\xi}\}\leq Ce^{-(l-vs)/\xi}$ in the former and $\min\{2,Ce^{-(l-vs)/\xi}\}\leq 2$ in the latter.
As a result,
\begin{align}
\|\hat{L}_i-\hat{L}_i^l\|\leq \frac{\sqrt{\gamma}}{\tauB}\int_0^{l/2v}ds\, Ce^{-2s/\tauB}e^{-(l-vs)/\xi}
\nonumber \\
+\frac{\sqrt{\gamma}}{\tauB}\int_{l/2v}^\infty ds\, 2e^{-2s/\tauB}.
\end{align}dqzzq
In the first term of the right hand side, $e^{-(l-vs)/\tauB}\leq e^{-l/2\tauB}$, and hence
\begin{align}
\|\hat{L}_i-\hat{L}_i^l\|&\leq\sqrt{\gamma}\left(\frac{C}{2}e^{-l/2\xi}+e^{-l/v\tauB}\right)
\nonumber \\
&\leq \sqrt{\gamma}\left(1+\frac{C}{2}\right)\exp\left(-\frac{l}{\max\{2\xi,v\tauB\}}\right),
\end{align}
which is the desired inequality.

\section{Full details of the error evaluation}
\label{sec:detail}

In the main text, we derive the error bound of $\epsilon_\mathrm{B}(t)$, i.e. the error induced by the Born approximation.
In the following, we explain full details of the evaluation of errors induced by the remaining approximations (the Markov approximation and the approximation associated with the memory-dressing transformation).

\subsection{Markov approximation}

In the Schr\"odinger picture, \cref{eq:xi_M} is expressed as
\begin{equation}
\hat{\xi}_\mathrm{M}(t)=\hat{\xi}_\mathrm{M}^{(1)}(t)+\hat{\xi}_\mathrm{M}^{(2)}(t),
\end{equation}
where
\begin{align}\label{eq:xi_M1}
\hat{\xi}_\mathrm{M}^{(1)}=\sum_{i\in\Lambda}\sum_{\nu=\pm 1}\nu\int_0^tds\, \Phi(\nu s)[\hat{A}_i,\mathcal{A}_i^\nu(-s)
\nonumber \\
\times e^{-i\HS t}(\rho^\mathrm{I}(t)-\rho^\mathrm{I}(t-s))e^{i\HS t}],
\end{align}
and
\begin{equation}\label{eq:xi_M2}
\hat{\xi}_\mathrm{M}^{(2)}(t)=\sum_{i\in\Lambda}\sum_{\nu=\pm 1}\nu\int_t^\infty ds\, \Phi(\nu s)[\hat{A}_i,\mathcal{A}_i^\nu(-s)\rho(t)].
\end{equation}

Let us start with the evaluation of the error caused by $\hat{\xi}_\mathrm{M}^{(2)}(t)$, which is given by
\begin{align}\label{eq:epsilon_M2}
&\epsilon_\mathrm{M}^{(2)}(t)=\int_0^t dt'\,\left|\TrS[\hat{O}_X^\mathrm{d}(t')\hat{\xi}_\mathrm{M}^{(2)}(t-t')]\right|
\nonumber \\
&\leq\sum_{i\in\Lambda}\sum_{\nu=\pm 1}\int_0^t dt'\int_{t-t'}^\infty ds\,\frac{\gamma}{\tauB}e^{-s/\tauB}\|\mathcal{A}_i^\nu(-s)[\hat{O}_X^\mathrm{d}(t'),\hat{A}_i]\|
\nonumber \\
&\leq 2\gamma\sum_{i\in\Lambda}\int_0^t dt'\, e^{-(t-t')/\tauB}\|[\hat{O}_X^\mathrm{d}(t'),\hat{A}_i]\|.
\end{align}

First, we consider a time-dependent system Hamiltonian $\HS(t)$ with no local conserved quantity.
By using \cref{eq:accelerated_decay_no} (accelerated dissipation) for $i$ with $d(X,i)\leq vt'$, and by using \cref{eq:LR} (the Lieb-Robinson bound) for $i$ with $d(X,i)>vt'$, we find that it is bounded as
\begin{equation}
\epsilon_\mathrm{M}^{(2)}(t)\leq\bar{\epsilon}_\mathrm{M}^{(2)}
\coloneqq \frac{4\zeta}{a}\tilde{\gamma}^{1/2}\tildetauB+4(|X|+C\xi)\tilde{\gamma}\tildetauB,
\end{equation}
which is of order $\tilde{\gamma}^{1/2}\tilde{\tau}_\mathrm{B}$ for small $\gamma$ or $\tauB$.

Next, we consider a static system Hamiltonian $\HS$ in which \cref{eq:accelerated_decay_diag,eq:accelerated_decay,eq:accelerated_decay_local} hold.
From \cref{eq:epsilon_M2}, we again decompose the summation over $i\in\Lambda$ into $d(X,i)\leq vt'$ and $d(X,i)>vt'$.
By using the Lieb-Robinson bound for the latter, we have
\begin{align}
\epsilon_\mathrm{M}^{(2)}(t)\leq 2\gamma&\int_0^tdt'\,e^{-(t-t')/\tauB}\sum_{i: d(X,i)\leq vt'}\|[\hat{O}_X^\mathrm{d}(t'),\hat{A}_i]\|
\nonumber \\
&+4C\xi\tilde{\gamma}\tildetauB.
\end{align}
We here decompose $\hat{O}_X^\mathrm{d}(t')=\hat{O}_\mathrm{diag}(t')+\sum_{x\in\Lambda}\hat{O}_\mathrm{off}^{(x)}(t')$.
From \cref{eq:accelerated_decay_diag}, the diagonal part vanishes in the thermodynamic limit.
As for the off-diagonal part, we can use \cref{eq:f}:
\begin{align}
\epsilon_\mathrm{M}^{(2)}(t)&\leq 2\gamma\int_0^tdt'\,e^{-\frac{t-t'}{\tauB}}\left(\zeta\tilde{\gamma}^{-1/2}e^{-a\tilde{\gamma}^{1/2}vt'}+\zeta'\tilde{\gamma}^{1/2}e^{-a'\gamma t'}\right)
\nonumber \\
&\qquad+4C\xi\tilde{\gamma}\tildetauB
\nonumber \\
&\leq 2\gamma\int_0^tdt'\, e^{-(t-t')/\tauB}\left(\zeta\tilde{\gamma}^{-1/2}+\zeta'\tilde{\gamma}^{1/2}\right)+4C\xi\tilde{\gamma}\tildetauB
\nonumber \\
&=2\zeta\tilde{\gamma}^{1/2}\tildetauB+2\zeta'\tilde{\gamma}^{3/2}\tildetauB+4C\xi\tilde{\gamma}\tildetauB,
\end{align}
which is again $O(\tilde{\gamma}^{1/2}\tildetauB)$.

The next task is to evaluate the error induced by $\hat{\xi}_\mathrm{M}^{(1)}(t)$.
In \cref{eq:xi_M1}, $\rho^\mathrm{I}(t)-\rho^\mathrm{I}(t-s)$ is written as
\begin{align}
&\rho^\mathrm{I}(t)-\rho^\mathrm{I}(t-s)
=\int_{t-s}^tds'\, \frac{d}{ds'}\rho^\mathrm{I}(s')
\nonumber \\
&=\int_{t-s}^t ds'\,\frac{d}{ds'}\TrB[\mathcal{U}_I(s',0)\rho(0)\otimes\rhoB]
\nonumber \\
&=-i\sum_{j\in\Lambda}\int_{t-s}^t ds'\, [\hat{A}_j(s'),\TrB(\hat{B}_j(s')\mathcal{U}_\mathrm{I}(s',0)\rho(0)\otimes\rhoB)].
\end{align}
By introducing 
\begin{equation}\label{eq:Psi_j}
\hat{\Psi}_j(t,t-s')\coloneqq -ie^{-i\HS t}\TrB[\hat{B}_j(s')\mathcal{U}_\mathrm{I}(s',0)\rho(0)\otimes\rhoB]e^{i\HS t},
\end{equation}
we obtain
\begin{align}\label{eq:rhot-rhos}
&e^{-i\HS t}(\rho^\mathrm{I}(t)-\rho^\mathrm{I}(t-s))e^{i\HS t}
\nonumber \\
&=\sum_{j\in\Lambda}\int_0^s ds'\, [\hat{A}_j(-s'),\hat{\Psi}_j(t,s')].
\end{align}
By using Wick's theorem, $\hat{\Psi}_j(t,t-s')$ is rewritten as
\begin{align}
&e^{i\HS t}\hat{\Psi}_j(t,t-s')e^{-i\HS t}=-i\int_0^{s'}ds''\, \Phi_j(s'-s'')\sum_{\nu=\pm 1}\nu
\nonumber \\
&\times\TrB[\mathcal{U}_\mathrm{I}(s',s'')\mathcal{A}_j^\nu(s'')\mathcal{U}_\mathrm{I}(s'',0)\rho(0)\otimes\rhoB],
\end{align}
and thus its trace norm is evaluated as
\begin{align}\label{eq:Psi_j_bound}
\|\hat{\Psi}_j(t,t-s')\|_1&\leq\int_0^{s'}ds''\, \frac{\gamma}{\tauB}e^{-(s'-s'')/\tauB}\times 2
\nonumber \\
&\leq 2\gamma.
\end{align}
By substituting \cref{eq:rhot-rhos} into \cref{eq:xi_M1}, we have
\begin{align}
\hat{\xi}_\mathrm{M}^{(1)}(t)=\sum_{ij\in\Lambda}\sum_{\nu=\pm 1}\nu\int_0^t ds\int_0^s ds'\, \Phi(\nu s)
\nonumber \\
\times [\hat{A}_i,\mathcal{A}_i^\nu(-s)[\hat{A}_j(-s'),\hat{\Psi}_j(t,s')]].
\end{align}

Thus, the error caused by $\hat{\xi}_\mathrm{M}^{(1)}(t)$ is given by
\begin{align}\label{eq:epsilon_M1_upper}
&\epsilon_\mathrm{M}^{(1)}(t)=\int_0^t dt'\left|\TrS[\hat{O}_X^\mathrm{d}(t')\hat{\xi}_\mathrm{M}^{(1)}(t-t')]\right|
\nonumber \\
&\leq \sum_{ij\in\Lambda}\int_0^\infty dt'\int_0^\infty ds\int_0^s ds'\, \frac{\gamma}{\tauB}e^{-s/\tauB}\sum_{\nu=\pm 1}
\nonumber \\
&\quad\times\left|\TrS\left\{[\mathcal{A}_i^\nu(-s)[\hat{O}_X^\mathrm{d}(t'),\hat{A}_i],\hat{A}_j(-s')]\hat{\Psi}_j(t,s')\right\}\right|
\nonumber \\
&\leq \frac{2\gamma^2}{\tauB}\sum_{ij\in\Lambda}\int_0^\infty dt'\int_0^\infty ds\int_0^s ds'\, e^{-s/\tauB}\sum_{\nu=\pm 1}
\nonumber \\
&\quad\times \|[\mathcal{A}_i^\nu(-s)[\hat{O}_X^\mathrm{d}(t'),\hat{A}_i],\hat{A}_j(-s')]\|
\nonumber \\
&\eqqcolon \bar{\epsilon}_\mathrm{M}^{(1)},
\end{align}
where we have used \cref{eq:Psi_j_bound}.
We notice that the expression of $\bar{\epsilon}_\mathrm{M}^{(1)}$ is similar to that of $\bar{\epsilon}_\mathrm{B}$ given in \cref{eq:epsilon_B_upper}, except for the presence of $\mathcal{A}_i^\nu(-s)$ in front of $[\hat{O}_X^\mathrm{d}(t'),\hat{A}_i]$.

Similarly to the analysis of the Born approximation, let us decompose the summation over $i$ and $j$ in \cref{eq:epsilon_M1_upper} into the three parts $\bar{\epsilon}_\mathrm{M}^{(1)}\leq \bar{\epsilon}_\mathrm{M}^\mathrm{(i)}+\bar{\epsilon}_\mathrm{M}^\mathrm{(ii)}+\bar{\epsilon}_\mathrm{M}^\mathrm{(iii)}$: (i) $i,j\in X(t)$, (ii) $i\in X^c(t)$ and $d(X,i)\geq d(X,j)$, and (iii)$j\in X^c(t)$ and $d(X,j)>d(X,i)$.
In the following, we evaluate each contribution separately.

\subsubsection*{(i) $i,j\in X(t)$}

In this case, we have
\begin{align}
\bar{\epsilon}_\mathrm{M}^\mathrm{(i)}\leq\frac{8\gamma^2}{\tauB}\int_0^\infty dt\int_0^\infty ds\int_0^s ds'\, e^{-s/\tauB}
\nonumber \\
\times \sum_{i,j\in X(t)}\|[\hat{O}_X^\mathrm{d}(t),\hat{A}_i]\|.
\end{align}
When the system Hamiltonian depends on time and has no local conserved quantity, we have already evaluated it in the Born approximation.
Therefore we immediately conclude $\bar{\epsilon}_\mathrm{M}^\mathrm{(i)}=O(\tilde{\gamma}^{1/2}\tilde{\tau}_\mathrm{B})$.

In a static system Hamiltonian that satisfies \cref{eq:accelerated_decay_diag,eq:accelerated_decay,eq:accelerated_decay_local}, we can simply evaluate $\bar{\epsilon}_\mathrm{M}^\mathrm{(i)}$ as follows.
By integrating  over $s$ and $s'$ and use $\sum_{j\in X(t)}=|X(t)|\leq |X|+2pvt$, we have
\begin{equation}\label{eq:ep_M(i)_static}
\bar{\epsilon}_\mathrm{M}^\mathrm{(i)}=8\gamma^2\tauB\int_0^\infty dt\, (|X|+2pvt)\sum_{i\in X(t)}\|[\hat{O}_X^\mathrm{d}(t),\hat{A}_i]\|.
\end{equation}
We now decompose $\hat{O}_X^\mathrm{d}(t)=\hat{O}_\mathrm{diag}(t)+\sum_{x\in\Lambda}\hat{O}_\mathrm{off}^{(x)}(t)$.
From \cref{eq:accelerated_decay_diag}, the contribution from $\hat{O}_\mathrm{diag}(t)$ vanishes in the thermodynamic limit.
Therefore we have
\begin{align}
\sum_{i\in X(t)}\|[\hat{O}_X^\mathrm{d}(t),\hat{A}_i]\|\leq \sum_{i,x\in\Lambda}\|[\hat{O}_\mathrm{off}^{(x)}(t),\hat{A}_i]\|
\nonumber \\
\approx\left(\zeta\tilde{\gamma}^{-1/2}e^{-a\tilde{\gamma}^{1/2}vt}+\zeta'\tilde{\gamma}^{1/2}e^{-a'\gamma t}\right),
\end{align}
where we have used \cref{eq:f} in the last line.
By substituting it into \cref{eq:ep_M(i)_static}, we obtain
\begin{equation}
\bar{\epsilon}_\mathrm{M}^\mathrm{(i)}\leq 16p\left(\frac{\zeta}{a^2}+\frac{\zeta'}{a'^2}\right)\tilde{\gamma}^{1/2}\tildetauB
+8|X|\left(\frac{\zeta}{a}\tilde{\gamma}+\frac{\zeta}{a'}\tilde{\gamma}^{3/2}\right)\tildetauB.
\end{equation}
It is $O(\tilde{\gamma}^{1/2}\tildetauB)$.

\subsubsection*{(ii) $i\in X^c(t)$ and $d(X,i)\geq d(X,j)$}

In this case, by using the simple bound $\|[\mathcal{A}_i^\nu(-s)[\hat{O}_X^\mathrm{d}(t),\hat{A}_i],\hat{A}_j(-s')]\|\leq 2\|[\hat{O}_X^\mathrm{d}(t),\hat{A}_i]\|$ and using the Lieb-Robinson bound, we have
\begin{equation}
\bar{\epsilon}_\mathrm{M}^\mathrm{(ii)}\leq 8C\gamma^2\tauB\int_0^\infty dt\, \sum_{i\in X(t)}\sum_{j:d(X,j)\leq d(X,i)}e^{-[d(X,i)-vt]/\xi}.
\end{equation}
This is identical to \cref{eq:epsilon_B_ii_upper}, and hence, following the same calculation, we conclude
\begin{equation}
\bar{\epsilon}_\mathrm{M}^\mathrm{(ii)}=O(\tilde{\gamma}^2\tilde{\tau}_\mathrm{B}).
\end{equation}

\subsubsection*{(iii) $j\in X^c(t)$ and $d(X,j)>d(X,i)$}

By using $\|\hat{A}_i\|=1$ and $[\hat{A}\hat{B},\hat{C}]=\hat{A}[\hat{B},\hat{C}]+[\hat{A},\hat{C}]\hat{B}$ for arbitrary operators $\hat{A}$, $\hat{B}$, and $\hat{C}$, we have
\begin{align}
&\|[\mathcal{A}_i^\nu(-s)[\hat{O}_X^\mathrm{d}(t),\hat{A}_i],\hat{A}_j(-s')]\|
\nonumber \\
&\leq \|[[\hat{O}_X^\mathrm{d}(t),\hat{A}_i],\hat{A}_j(-s')]\|
\nonumber \\
&\quad+\|[\hat{A}_i(-s),\hat{A}_j(-s')]\|\cdot\|[\hat{O}_X^\mathrm{d}(t),\hat{A}_i]\|.
\end{align}
By using this inequality, we obtain
\begin{equation}
\bar{\epsilon}_\mathrm{M}^\mathrm{(iii)}\leq \bar{\epsilon}_\mathrm{M}^\mathrm{(iii-1)}+\bar{\epsilon}_\mathrm{M}^\mathrm{(iii-2)},
\end{equation}
where
\begin{align}
\bar{\epsilon}_\mathrm{M}^\mathrm{(iii-1)}&=\frac{4\gamma^2}{\tauB}\int_0^\infty dt\int_0^\infty ds\int_0^s ds'\, e^{-s/\tauB}
\nonumber \\
&\quad\times\sum_{j\in X^c(t)}\sum_{i:d(X,i)<d(X,j)}
\|[[\hat{O}_X^\mathrm{d}(t),\hat{A}_i],\hat{A}_j(-s')]\|,
\end{align}
and
\begin{align}\label{eq:epsilon_M_iii-2}
&\bar{\epsilon}_\mathrm{M}^\mathrm{(iii-2)}=\frac{4\gamma^2}{\tauB}\int_0^\infty dt\int_0^\infty ds\int_0^s ds'\, e^{-s/\tauB}
\nonumber \\
&\times\sum_{j\in X^c(t)}\sum_{i:d(X,i)<d(X,j)}\|[\hat{A}_i(-s'),\hat{A}_j]\|\cdot\|[\hat{O}_X^\mathrm{d}(t),\hat{A}_i]\|.
\end{align}
We find that $\bar{\epsilon}_\mathrm{M}^\mathrm{(iii-1)}$ is identical to $\bar{\epsilon}_\mathrm{B}^\mathrm{(iii)}$, and hence $\bar{\epsilon}_\mathrm{M}^\mathrm{(iii-1)}=O(\tilde{\gamma}^2\tilde{\tau}_\mathrm{B})$.

Thus, the remaining problem is to evaluate $\bar{\epsilon}_\mathrm{M}^\mathrm{(iii-2)}$.
The product $\|[\hat{A}_i(-s'),\hat{A}_j]\|\cdot\|[\hat{O}_X^\mathrm{d}(t),\hat{A}_i]\|$ in \cref{eq:epsilon_M_iii-2} is evaluated by using the Lieb-Robinson bound.
We bound it as follows: when $i\in X_j$,
\begin{align}
&\|[\hat{A}_i(-s'),\hat{A}_j]\|\cdot\|[\hat{O}_X^\mathrm{d}(t),\hat{A}_i]\|\leq
2\min\{2, Ce^{-[d(i,j)-vs']/\xi}\}
\nonumber \\
&\leq 2\min\left\{2, C\exp\left[-\frac{1}{\xi}\left(\frac{d(X,j)}{p}-vs'\right)\right]\right\},
\end{align}
where we used $d(i,j)\geq d(X,j)-d(X,i)\geq d(X,j)/p$, and when $i\in X_j^c$,
\begin{align}
&\|[\hat{A}_i(-s'),\hat{A}_j]\|\cdot\|[\hat{O}_X^\mathrm{d}(t),\hat{A}_i]\|
\leq 2Ce^{-[d(X,i)-vt]/\xi}
\nonumber \\
&\leq 2C\exp\left[-\frac{1}{\xi}\left(\frac{p-1}{p}d(X,j)-vt\right)\right],
\end{align}
where we used $d(X,i)>d(X,j)(p-1)/p$ (recall the definition of $X_j$, see \cref{eq:Xj}).
By substituting these inequalities into \cref{eq:epsilon_M_iii-2}, after calculations, we finally obtain
\begin{align}
&\bar{\epsilon}_\mathrm{M}^\mathrm{(iii-2)}\leq
\tilde{\gamma}^2\tilde{\tau}_\mathrm{B}\cdot 16Cp\xi\left[\frac{p(2p-3)}{(p-1)^2(p-2)^2}\xi^2\right.
\nonumber \\
&\left.\vphantom{\frac{1}{1}}+|X|(\xi+\tilde{\tau}_\mathrm{B})+4(p-1)(\xi^2+\xi\tilde{\tau}_\mathrm{B}+\tilde{\tau}_\mathrm{B}^2\right]
\nonumber \\
&+\tilde{\gamma}^2\tilde{\tau}_\mathrm{B}^3\cdot 32p[|X|+4(p-1)\tilde{\tau}_\mathrm{B}].
\end{align}
From this expression, we find that $\epsilon_\mathrm{M}^\mathrm{(iii-2)}=O(\tilde{\gamma}^2\tilde{\tau}_\mathrm{B})$.

In conclusion, we find $\bar{\epsilon}_\mathrm{M}^\mathrm{(iii)}\leq \bar{\epsilon}_\mathrm{M}^\mathrm{(iii-1)}+\bar{\epsilon}_\mathrm{M}^\mathrm{(iii-2)}=O(\tilde{\gamma}^2\tilde{\tau}_\mathrm{B})$.

\subsection{Memory-dressing transformation}
\label{sec:mod}

By looking at \cref{eq:epsilon_mod}, we see that $\epsilon_\mathrm{md}(t)$ is composed of four parts:
\begin{equation}\left\{
\begin{aligned}
&\epsilon_\mathrm{md}^{(1)}(t)=\int_0^t dt'\, \left|\TrS[\hat{O}_X^\mathrm{d}(t')\hat{\xi}_1(t-t')]\right|,
\\ 
&\epsilon_\mathrm{md}^{(2)}(t)=\int_0^t dt'\, \left|\TrS[\hat{O}_X^\mathrm{d}(t')\hat{\xi}_2(t-t')]\right|,
\\ 
&\epsilon_\mathrm{md}^{(3)}(t)=\left|\TrS[\hat{O}_X(\rho(t)-\tilde{\rho}(t))]\right|,
\\ 
&\epsilon_\mathrm{md}^{(4)}(t)=\left|\TrS[\hat{O}_X^\mathrm{d}(t)(\rho(0)-\tilde{\rho}(0))]\right|.
\end{aligned}
\right.
\end{equation}
We evaluate each contribution.

By going back to the Schr\"odinger picture, $\hat{\xi}_1(t)=e^{-i\HS t}\hat{\xi}_1^\mathrm{I}(t)e^{i\HS t}$ is given by
\begin{equation}\label{eq:xi_1_Sch}
\hat{\xi}_1(t)=\mathcal{M}(0)\left[e^{-i\HS t}\frac{\pd\rho^\mathrm{I}(t)}{\pd t}e^{i\HS t}\right],
\end{equation}
where $\mathcal{M}(t)$ is defined in \cref{eq:M}, which is expressed as
\begin{equation}\label{eq:M(0)}
\mathcal{M}(0)=\sum_{\nu=\pm 1}\int_0^\infty ds_1\int_{-\infty}^0ds_2\int_{-\infty}^\infty ds_3\, \nu\mathcal{F}(\nu s_1,\nu s_2,s_3).
\end{equation}
By using $\hat{\Psi}_j(t,s)$ defined in \cref{eq:Psi_j}, we can write
\begin{equation}\label{eq:drhoIdt}
e^{-i\HS t}\frac{\pd\rho^\mathrm{I}(t)}{\pd t}e^{i\HS t}
=\sum_{j\in\Lambda}[\hat{A}_j,\hat{\Psi}_j(t,0)].
\end{equation}
By substituting \cref{eq:M(0),eq:drhoIdt} into \cref{eq:xi_1_Sch}, $\epsilon_\mathrm{md}^{(1)}(t)$ is bounded from above as
\begin{align}
\epsilon_\mathrm{md}^{(1)}(t)\leq 2\sum_{\nu=\pm 1}&\int_0^t dt'\int_0^\infty ds_1\int_{-\infty}^0 ds_2\int_{-\infty}^{\nu s_1}ds_3
\nonumber \\
&\times |g_i(\nu s_1-\nu s_2)|\cdot |g_i(\nu s_2-s_3)| 
\nonumber \\
&\times G(t',\nu s_1,s_3)\|\Psi_j(t-t',0)\|_1,
\end{align}
where
\begin{equation}
G(t,s,s')=\sum_{ij\in\Lambda}\|[\hat{A}_i(s')[\hat{O}_X^d(t),\hat{A}_i(s)],\hat{A}_j]\|.
\end{equation}
By using \cref{eq:ineq_g,eq:Psi_j_bound}, we obtain
\begin{align}\label{eq:bar_epsilon_mod1}
\epsilon_\mathrm{md}^{(1)}(t)&\leq\frac{\gamma^2}{\tauB^2}\sum_{\nu=\pm 1}
\int_0^\infty dt'\int_0^\infty ds_1\int_{-\infty}^0 ds_2\int_{-\infty}^{\nu s_1}ds_3
\nonumber \\
&\quad \times e^{-2(s_1-s_2)/\tauB}e^{-2|\nu s_2-s_3|/\tauB}G(t',\nu s_1,s_3)
\nonumber \\
&\eqqcolon \bar{\epsilon}_\mathrm{md}^{(1)}.
\end{align}
By following similar calculations as in the previous subsections, an upper bound on $\bar{\epsilon}_\mathrm{md}^{(1)}$ is evaluated in \cref{sec:mod1}, where it turns out that 
\begin{equation}
\bar{\epsilon}_\mathrm{md}^{(1)}\lesssim \tilde{\gamma}^{1/2}\tilde{\tau}_\mathrm{B}
\end{equation}
for small $\gamma$ or small $\tauB$.

Next, we consider $\epsilon_\mathrm{md}^{(2)}(t)$.
By using $\hat{\xi}_2(t)=-\tilde{\mathcal{L}}\mathcal{M}(0)\rho(t)$, we obtain
\begin{align}
\epsilon_\mathrm{md}^{(2)}(t)&=\int_0^t dt'\,\left|\TrS\left[\left(\mathcal{M}^\dagger(0)\tilde{\mathcal{L}}^\dagger\hat{O}_X^\mathrm{d}(t')\right)\rho(t)\right]\right|
\nonumber \\
&\leq \int_0^t dt'\,\|\mathcal{M}^\dagger(0)\tilde{\mathcal{L}}^\dagger\hat{O}_X^\mathrm{d}(t')\|,
\end{align}
where $\mathcal{M}^\dagger(0)$ is the adjoint of $\mathcal{M}(0)$: $\TrS[\hat{X}_1\mathcal{M}(0)\hat{X}_2]=\TrS[(\mathcal{M}^\dagger(0)\hat{X}_1)\hat{X}_2]$ for arbitrary operators $\hat{X}_1$ and $\hat{X}_2$.
By using \cref{eq:ineq_g,eq:F,eq:M} combined with the inequality $\int_{-\infty}^s ds_3\, e^{-2|s_2-s_3|/\tauB}\leq\tauB$ for an arbitrary $s\in\mathbb{R}$, we obtain the following simple bound on $\epsilon_\mathrm{md}^{(2)}(t)$:
\begin{align}\label{eq:bar_epsilon_mod2}
\epsilon_\mathrm{md}^{(2)}(t)&\leq\frac{\gamma}{4}\sum_{\nu=\pm 1}\sum_{i,j\in\Lambda}\int_0^\infty dt'\int_0^\infty ds\, e^{-2s/\tauB}
\nonumber \\
&\qquad\times\|[\tilde{\mathcal{L}}_i^\dagger\hat{O}_X^\mathrm{d}(t'),\hat{A}_j(\nu s)]\|
\nonumber \\
&\eqqcolon \bar{\epsilon}_\mathrm{md}^{(2)}.
\end{align}
where $\tilde{\mathcal{L}}_i^\dagger=\mathcal{L}_{\Delta,i}^\dagger+\mathcal{D}_i^\dagger$.

The evaluation of $\bar{\epsilon}_\mathrm{md}^{(2)}$ is provided in \cref{sec:mod2}.
It turns out that the contribution from $i,j\in X(t)$ is dominant, and we again obtain
\begin{equation}
\bar{\epsilon}_\mathrm{md}^{(2)}\lesssim \tilde{\gamma}^{1/2}\tilde{\tau}_\mathrm{B}. 
\end{equation}

Next, let us consider $\epsilon_\mathrm{md}^{(3)}(t)$.
It is written as
\begin{equation}\label{eq:ep_mod3}
\epsilon_\mathrm{md}^{(3)}(t)=\left|\TrS[\hat{O}_X^\mathrm{d}(t)\mathcal{M}(0)\rho(t)]\right|
\leq \|\mathcal{M}^\dagger(0)\hat{O}_X^\mathrm{d}(t)\|.
\end{equation}
By using \cref{eq:M(0),eq:ineq_g}, we have
\begin{align}\label{eq:ep_mod3_upper}
&\epsilon_\mathrm{md}^{(3)}(t)\leq \frac{\gamma}{2\tauB^2}\sum_{\nu=\pm 1}\sum_{i\in\Lambda}\int_0^\infty ds_1\int_{-\infty}^0ds_2\int_{-\infty}^{\nu s_1} ds_3
\nonumber \\
&\times e^{-2(s_1-s_2)/\tauB}e^{-2|\nu s_2-s_3|/\tauB}\|[\hat{O}_X^\mathrm{d}(t),\hat{A}_i(\nu s_1)]\|.\end{align}
Let us evaluate it for $i\in X(t)$ and $i\in X^c(t)$.
For $i\in X(t)$, the calculation is almost identical to that for $\bar{\epsilon}_\mathrm{md}^{(1)}$ given in \cref{sec:mod1}.
It turns out that the contribution from $i\in X(t)$ in the right hand side of \cref{eq:ep_mod3_upper} is $O(\tilde{\gamma}^{1/2}\tildetauB)$.
For $i\in X^c(t)$, we use the Lieb-Robinson bound, e.g. \cref{eq:LR}:
\begin{align}
&\frac{\gamma}{2\tauB^2}\sum_{\nu=\pm 1}\sum_{i\in X^c(t)}\int_0^\infty ds_1\int_{-\infty}^0ds_2\int_{-\infty}^{\nu s_1}ds_3
\nonumber \\
&\times e^{-2(s_1-s_2)/\tauB}e^{-2|\nu s_2-s_3|/\tauB}\min\{2,Ce^{-[d(X,i)-v(t+s_1)]/\xi}\}.
\end{align}
For $s_1\leq(p-1)d(X,i)/pv$, we use
\begin{align}
\min\{2,Ce^{-[d(X,i)-v(t+s_1)]/\xi}\}&\leq Ce^{-[d(X,i)-v(t+s_1)]/\xi}
\nonumber \\
&\leq Ce^{-\left[\frac{d(X,i)}{p}-vt\right]/\xi},
\end{align}
and for $s_1>(p-1)d(X,i)/pv$, we use 
\begin{equation}
\min\{2,Ce^{-[d(X,i)-v(t+s_1)]/\xi}\}\leq 2.
\end{equation}
We then obtain an upper bound
\begin{equation}\label{eq:ep_mod3_Xc}
\frac{Cp\xi}{4}\tilde{\gamma}\tilde{\tau}_\mathrm{B}+\frac{p}{4(p-1)}e^{-2(p-1)t/\tauB}\tilde{\gamma}\tilde{\tau}_\mathrm{B}^2
\end{equation}
from $i\in X^c(t)$ in \cref{eq:ep_mod3_upper}.

An upper bound on $\epsilon_\mathrm{md}^{(3)}(t)$ is given by the sum of the two contributions, from $i\in X(t)$ and $i\in X^c(t)$.
We therefore find that $\epsilon_\mathrm{md}^{(3)}(t)$ behaves as
\begin{equation}
\epsilon_\mathrm{md}^{(3)}(t)\lesssim \tilde{\gamma}^{1/2}\tilde{\tau}_\mathrm{B}
\end{equation}
for small $\gamma$ or small $\tauB$.

Finally, we consider $\epsilon_\mathrm{md}^{(4)}(t)$.
We find
\begin{equation}
\epsilon_\mathrm{md}^{(4)}(t)=\left|\TrS[\hat{O}_X^\mathrm{d}(t)\mathcal{M}(0)\rho(0)]\right|
\leq\|\mathcal{M}^\dagger(0)\hat{O}_X^\mathrm{d}(t)\|,
\end{equation}
which is identical to the right hand side of \cref{eq:ep_mod3}.
We can therefore conclude that $\epsilon_\mathrm{md}^{(4)}(t)$ behaves as
\begin{equation}
\epsilon_\mathrm{md}^{(4)}(t)\lesssim\tilde{\gamma}^{1/2}\tilde{\tau}_\mathrm{B}
\end{equation}
for small $\gamma$ or small $\tauB$.

\subsection{Evaluation of $\bar{\epsilon}_\mathrm{md}^{(1)}$}
\label{sec:mod1}

We evaluate an upper bound on 
\begin{equation}
G(t,s,s')=\sum_{ij\in\Lambda}\|[\hat{A}_i(s')[\hat{O}_X^\mathrm{d}(t),\hat{A}_i(s)],\hat{A}_j]\|,
\end{equation}
which is used to evaluate $\bar{\epsilon}_\mathrm{md}^{(1)}$ given by \cref{eq:bar_epsilon_mod1}.
For a fixed $t$, the sum over $i$ and $j$ is decomposed into the three parts: (i) $i,j\in X(t)$, (ii) $i\in X^c(t)$ and $j$ with $d(X,j)\leq d(X,i)$, and (iii) $j\in X^c(t)$ and $i$ with $d(X,i)\leq d(X,j)$.
Each contribution is denoted by $G^\mathrm{(i)}(t,s,s')$, $G^\mathrm{(ii)}(t,s,s')$, and $G^\mathrm{(iii)}(t,s,s)$, respectively.

\subsubsection*{(i) $i,j\in X(t)$}
First, let us consider a time-dependent system Hamiltonian with no local conserved quantity.
In this case, we can simply evaluate $G^\mathrm{(i)}(t,s,s')$ as
\begin{equation}\label{eq:G(i)0}
G^\mathrm{(i)}(t,s,s')\leq 2\sum_{i,j\in X(t)}\|[\hat{O}_X^\mathrm{d}(t),\hat{A}_i(s)]\|.
\end{equation}
Here, we use the acceleration of dissipation, i.e. \cref{eq:accelerated_decay_no}:
\begin{align}\label{eq:G(i)}
G^\mathrm{(i)}(t,s,s')\leq 2\zeta(|X|+2pvt)^2e^{-a\tilde{\gamma}^{1/2}vt}.
\end{align}

By using \cref{eq:G(i)}, we can evaluate the contribution to $\bar{\epsilon}_\mathrm{md}^{(1)}$ from $i,j\in X(t)$, which is denoted by $\bar{\epsilon}_\mathrm{md}^\mathrm{(1-i)}$.
After an explicit calculation, we obtain
\begin{align}
&\sum_{\nu=\pm 1}\int_0^\infty ds_1\int_{-\infty}^0 ds_2\int_{-\infty}^{\nu s_1}ds_3\, e^{-2(s_1-s_2)/\tauB}e^{-2|\nu s_2-s_3|/\tauB}
\nonumber \\
&=\frac{\tauB^3}{4}.
\end{align}
By using it and \cref{eq:G(i)} in \cref{eq:bar_epsilon_mod1}, we find that 
\begin{equation}
\bar{\epsilon}_\mathrm{md}^\mathrm{(1-i)}\leq \left(\frac{4p^2}{a^3}\tilde{\gamma}^{1/2}
+\frac{2p|X|}{a^2}\tilde{\gamma}
+\frac{|X|^2}{2a}\tilde{\gamma}^{3/2}\right)\tildetauB\zeta,
\end{equation}
which is of $O(\tilde{\gamma}^{1/2}\tildetauB)$. 

Next, we consider a static Hamiltonian satisfying \cref{eq:accelerated_decay_diag,eq:accelerated_decay,eq:accelerated_decay_local}.
By decomposing $\hat{O}_X^\mathrm{d}(t)$ into the diagonal part and the off-diagonal part as $\hat{O}_X^\mathrm{d}(t)=\hat{O}_\mathrm{diag}(t)+\sum_{x\in\Lambda}\hat{O}_\mathrm{off}^{(x)}(t)$, we can ignore the diagonal part in \cref{eq:G(i)0} in the thermodynamic limit by using \cref{eq:accelerated_decay_diag}.
We therefore have
\begin{align}
&G^\mathrm{(i)}(t,s,s')\leq 2|X(t)|\sum_{i,x\in\Lambda}\|[\hat{O}_\mathrm{off}^{(x)}(t),\hat{A}_i(s)]\|
\nonumber \\
&=2|X(t)|\sum_{x\in\Lambda}\left(\sum_{i:d(R_x,i)<v|s|}\|[\hat{O}_\mathrm{off}^{(x)}(t),\hat{A}_i(s)]\|
\right.\nonumber \\
&\left.\quad +\sum_{i:d(R_x,i)\geq v|s|}C\|\hat{O}_\mathrm{off}^{(x)}\|e^{-[d(R_x,i)-v|s|]/\xi}\right)
\nonumber \\
&\leq 2|X(t)|\left[4(\ell_\gamma+v|s|)+C\xi\right]\sum_{x\in\Lambda}\|\hat{O}_\mathrm{off}^{(x)}(t)\|,
\end{align}
where $R_x=\{i\in\Lambda: d(i,x)<\ell_\gamma\}$ and \cref{eq:accelerated_decay_local} is used in the second line (i.e. when $d(R_x,i)\geq v|s|$, we can use the Lieb-Robinson bound: $\|[\hat{O}_\mathrm{off}^{(x)}(t),\hat{A}_i(s)]\|\leq C\|\hat{O}_\mathrm{off}^{(x)}\|e^{-[d(R_x,i)-v|s|]/\xi}$).
By using \cref{eq:accelerated_decay}, we have
\begin{align}
G^\mathrm{(i)}(t,s,s')\leq 2|X(t)|[4(\ell_\gamma+v|s|)+C\xi]
\nonumber \\
\times \left(\zeta e^{-a\tilde{\gamma}^{1/2}vt}+\zeta'\tilde{\gamma}e^{-a'\gamma t}\right).
\end{align}
By substituting it into \cref{eq:bar_epsilon_mod1} and carrying out the integral, we conclude
\begin{equation}
\bar{\epsilon}_\mathrm{md}^{(1-i)}=O(\tilde{\gamma}^{1/2}\tildetauB).
\end{equation}

\subsubsection*{(ii) $i\in X^c(t)$ and $d(X,j)\leq d(X,i)$}
In this case, $G^\mathrm{(ii)}(t,s,s')$ is evaluated as
\begin{equation}
G^\mathrm{(ii)}(t,s,s')\leq 2\sum_{i\in X^c(t)}\sum_{j:d(X,j)\leq d(X,i)}\|[\hat{O}_X^\mathrm{d}(t),\hat{A}_i(s)]\|.
\end{equation}
By putting $d(X,i)=r$, we have $\sum_{i\in X^c(t)}\approx 2\int_{pvt}^\infty dr$ and $\sum_{j: d(X,j)\leq d(X,i)}=|X|+2r$.
By using the Lieb-Robinson bound, we obtain
\begin{equation}
G^\mathrm{(ii)}(t,s,s')\leq 4\int_{pvt}^\infty dr\, (|X|+2r)\min\{2,Ce^{-[r-v(t+|s|)]/\xi}.
\end{equation}
As for $\min\{2, Ce^{-[r-v(t+|s|)]/\xi}$, when $|s|\leq (p-2)t$, we choose $Ce^{-[r-v(t+|s|)]/\xi}$.
When $|s|>(p-2)t$, we choose $Ce^{-[r-v(t+|s|)]/\xi}$ for $r>pv|s|/(p-2)$, and 2 for $r\leq pv|s|/(p-2)$.
As a result, we obtain the following upper bound on $G^\mathrm{(ii)}(t,s,s')$:
\begin{align}
&4I[|s|\leq (p-2)t]C\int_{pvt}^\infty dr\, (|X|+2r)e^{-[r-v(t+|s|)]/\xi}
\nonumber \\
+&4I[|s|>(p-2)t]\left[\int_{pvt}^{\frac{p}{p-2}v|s|}dr\, 2(|X|+2r)\right.
\nonumber \\
&\left.+\int_{\frac{p}{p-2}v|s|}^\infty dr\, (|X|+2r)Ce^{-[r-v(t+|s|)]/\xi}\right].
\end{align}
By performing the integration over $r$, we obtain
\begin{widetext}
\begin{align}
&G^\mathrm{(ii)}(t,s,s')\leq I[|s|\leq(p-2)t]\cdot 4C\xi(|X|+2\xi+2pvt)e^{-vt/\xi}
\nonumber \\
&+I[|s|>(p-2)t]\left\{\frac{2p}{p-2}v|X|[|s|-(p-2)t]+2\left(\frac{p}{p-2}\right)^2v^2[|s|^2-(p-2)^2t^2]
+C\xi\left(|X|+2\xi+\frac{2p}{p-2}v|s|\right)\right\}.
\end{align}
\end{widetext}
By substituting it into \cref{eq:bar_epsilon_mod1}, after some calculations, we finally obtain $\bar{\epsilon}_\mathrm{md}^\mathrm{(1-ii)}$, which corresponds to the contribution from $i,j$ with $i\in X^c(t)$ and $d(X,j)\leq d(X,i)$, as follows:
\begin{align}
&\bar{\epsilon}_\mathrm{md}^\mathrm{(1-ii)}\leq \tilde{\gamma}^2\tilde{\tau}_\mathrm{B}C\xi^2[|X|+2(p+1)\xi]
\nonumber \\
&+\frac{\tilde{\gamma}^2\tilde{\tau}_\mathrm{B}^2}{8(p-2)^3}[C\xi(p-2)^2(|X|+2\xi)
\nonumber \\
&\qquad+p(p-2)(|X|+2C\xi)\tilde{\tau}_\mathrm{B}+2p^2\tilde{\tau}_\mathrm{B}^2].
\end{align}
For small $\gamma$ or small $\tauB$, we find $\epsilon_\mathrm{md}^\mathrm{(1-ii)}\lesssim\tilde{\gamma}^2\tilde{\tau}_\mathrm{B}$.

\subsubsection*{(iii) $j\in X^c(t)$ and $d(X,i)\leq d(X,j)$}
By using $[\hat{A}\hat{B},\hat{C}]=\hat{A}[\hat{B},\hat{C}]+[\hat{A},\hat{C}]\hat{B}$ for arbitrary operators $A$, $B$, and $C$, we have
\begin{align}\label{eq:G(iii)}
&G^\mathrm{(iii)}(t,s,s')\leq\sum_{j\in X^c(t)}\sum_{i:d(X,i)\leq d(X,j)}
\nonumber \\
&\left(\|[[\hat{O}_X^\mathrm{d}(t),\hat{A}_i(s)],\hat{A}_j]\|+\|[\hat{A}_i(s'),\hat{A}_j]\|\cdot\|[\hat{O}_X^\mathrm{d}(t),\hat{A}_i(s)]\|\right).
\end{align}

By defining $\hat{O}_X^\mathrm{d}(t,s)\coloneqq e^{i\HS s}\hat{O}_X^\mathrm{d}(t)e^{-i\HS s}$, we have
\begin{align}\label{eq:G(iii)_1}
&\|[[\hat{O}_X^\mathrm{d}(t),\hat{A}_i(s)],\hat{A}_j]\|
=\|[[\hat{O}_X^\mathrm{d}(t,-s),\hat{A}_i],\hat{A}_j(-s)]\|
\nonumber \\
&\leq\|[[\hat{O}_{X,j}(t,-s),\hat{A}_i],\hat{A}_j(-s)]\|
\nonumber \\
&\quad+4\|\hat{O}_X^\mathrm{d}(t,-s)-\hat{O}_{X,j}^\mathrm{d}(t,-s)\|,
\end{align}
where $\hat{O}_{X,j}(t,s)$ is the operator that is obtained by localizing $\hat{O}_X^\mathrm{d}(t,s)$ to the region $X_j$ defined by \cref{eq:Xj}; see \cref{eq:O_Xj}.
By using the Lieb-Robinson bound, we have
\begin{align}\label{eq:G(iii)_1-1}
&\|[[\hat{O}_{X,j}(t,-s),\hat{A}_i],\hat{A}_j(-s)]\|
\nonumber \\
&\leq I[i\in X_j]\min\left\{2,Ce^{-\left[\frac{d(X,j)}{p}-v|s|\right]/\xi}\right\},
\end{align}
where we have used $d(X_j,j)\geq d(X,j)/p$, and
\begin{align}\label{eq:G(iii)_1-2}
&\|\hat{O}_X^\mathrm{d}(t,-s)-\hat{O}_{X,j}^\mathrm{d}(t,-s)\|
\nonumber \\
&\leq \min\{2,C|X|e^{-\left[\frac{p-1}{p}d(X,j)-v(t+|s|)\right]/\xi}\}.
\end{align}

As for $\|[\hat{A}_i(s'),\hat{A}_j]\|\cdot\|[\hat{O}_X^\mathrm{d}(t),\hat{A}_i(s)]\|$ in \cref{eq:G(iii)}, the LR bound is used for $\|[\hat{A}_i(s'),\hat{A}_j]\|$ when $i\in X_j$, and for $\|[\hat{O}_X^\mathrm{d}(t),\hat{A}_i(s)]\|$ when $i\in X_j^c$.
We then have
\begin{align}\label{eq:G(iii)_2}
&\|[\hat{A}_i(s'),\hat{A}_j]\|\cdot\|[\hat{O}_X^\mathrm{d}(t),\hat{A}_i(s)]\|
\nonumber \\
&\leq 2I[i\in X_j]\min\{2,Ce^{-\left[\frac{d(X,j)}{p}-v|s'|\right]/\xi}\}
\nonumber \\
&\quad+2I[i\in X_j^c]\min\{2,Ce^{-\left[\frac{p-1}{p}d(X,j)-v(t+|s|)\right]/\xi}\},
\end{align}
where we used $d(i,j)\geq d(X,j)/p$ for $i\in X_j$ and $d(X,i)\geq [(p-1)/p]d(X,j)$ for $i\in X_j^c$.

By substituting above results, i.e. \cref{eq:G(iii)_1,eq:G(iii)_1-1,eq:G(iii)_1-2,eq:G(iii)_2}, into \cref{eq:G(iii)}, we obtain an upper bound on $G^\mathrm{(iii)}(t,s,s')$ and $\bar{\epsilon}_\mathrm{md}^\mathrm{(1-iii)}$ by using \cref{eq:bar_epsilon_mod1}.
Everything is finite, and we find
\begin{equation}
\bar{\epsilon}_\mathrm{md}^\mathrm{(1-iii)}\lesssim \tilde{\gamma}^2\tilde{\tau}_\mathrm{B}
\end{equation}
for small $\gamma$ or small $\tauB$.

\subsection{Evaluation of $\bar{\epsilon}_\mathrm{md}^{(2)}$}
\label{sec:mod2}

We evaluate $\bar{\epsilon}_\mathrm{md}^{(2)}$, which is defined by \cref{eq:bar_epsilon_mod2}.
As before, we decompose the sum over $i,j\in\Lambda$ into the three parts: (i) $i,j\in X(t)$, (ii) $i\in X^c(t)$ and $d(X,j)\leq d(X,i)$, and (iii) $j\in X^c(t)$ and $d(X,j)\leq d(X,i)$.
Each contribution is denoted by $\bar{\epsilon}_\mathrm{md}^\mathrm{(2-i)}$, $\bar{\epsilon}_\mathrm{md}^\mathrm{(2-ii)}$, and $\bar{\epsilon}_\mathrm{md}^\mathrm{(2-iii)}$.

\subsubsection*{(i) $i,j\in X(t)$}
From \cref{eq:bar_epsilon_mod2}, we have
\begin{align}\label{eq:bar_epsilon_mod2-i}
\bar{\epsilon}_\mathrm{md}^\mathrm{(2-i)}&\leq \gamma\int_0^\infty dt\int_0^\infty ds\, e^{-2s/\tauB}\sum_{i,j\in X(t)}\|\mathcal{L}_i^\dagger\hat{O}_X^\mathrm{d}(t)\|
\nonumber \\
&=\frac{\gamma\tauB}{2}\int_0^\infty dt\sum_{i\in X(t)}(|X|+2pvt)\|\mathcal{L}_i^\dagger\hat{O}_X^\mathrm{d}(t)\|.
\end{align}
From the expression
\begin{equation}
\mathcal{L}_i^\dagger\hat{O}_X^\mathrm{d}(t)=i[\hat{\Delta}_i,\hat{O}_X^\mathrm{d}(t)]+\hat{L}_i^\dagger\hat{O}_X^\mathrm{d}(t)\hat{L}_i^\dagger-\frac{1}{2}\{\hat{L}_i^\dagger\hat{L}_i,\hat{O}_X^\mathrm{d}(t)\},
\end{equation}
we obtain
\begin{align}
\|\mathcal{L}_i^\dagger\hat{O}_X^\mathrm{d}(t)]\|\leq\|[\hat{O}_X^\mathrm{d}(t),\hat{\Delta}_i]\|+\frac{1}{2}\|\hat{L}_i^\dagger\|\cdot\|[\hat{O}_X^\mathrm{d}(t),\hat{L}_i]\|
\nonumber \\
+\frac{1}{2}\|\hat{L}_i\|\cdot\|[\hat{O}_X^\mathrm{d}(t),\hat{L}_i^\dagger]\|.
\end{align}

First, we consider a time-dependent system Hamiltonian without local conserved quantity.
By using $\|\hat{L}_i\|=\|\hat{L}_i^\dagger\|\leq\sqrt{\gamma}/2$ and $\|\hat{\Delta}_i\|\leq\gamma/8$, and using the accelerated dissipation~(\ref{eq:accelerated_decay_no}), we obtain
\begin{align}
\bar{\epsilon}_\mathrm{md}^\mathrm{(2-i)}\leq\frac{3}{16}\gamma^2\int_0^\infty dt\, (|X|+2pvt)^2\zeta e^{-a\tilde{\gamma}^{1/2}vt},
\end{align}
which is of $O(\tilde{\gamma}^{1/2}\tilde{\tau}_\mathrm{B})$ for small $\gamma$ or small $\tauB$.

Next, we consider a static system Hamiltonian that satisfies \cref{eq:accelerated_decay_diag,eq:accelerated_decay,eq:accelerated_decay_local}.
From \cref{eq:accelerated_decay_diag}, the contribution from the diagonal part $\hat{O}_\mathrm{diag}(t)$ is ignorable in $\sum_{i\in X(t)}\|\mathcal{L}_i^\dagger\hat{O}_X^\mathrm{d}(t)\|$.
We therefore have, in the thermodynamic limit,
\begin{align}
&\sum_{i\in X(t)}\|\mathcal{L}_i^\dagger\hat{O}_X^\mathrm{d}(t)\|\leq\sum_{x,i\in\Lambda}\|\mathcal{L}_i^\dagger\hat{O}_\mathrm{off}^{(x)}(t)\|
\nonumber \\
&\leq \sum_{x,i\in\Lambda}\left(\|[\hat{O}_\mathrm{off}^{(x)}(t),\hat{\Delta}_i]\|+\frac{\sqrt{\gamma}}{4}\|[\hat{O}_\mathrm{off}^{(x)}(t),\hat{L}_i]\|
\right.\nonumber \\
&\quad \left.+\frac{\sqrt{\gamma}}{4}\|[\hat{O}_\mathrm{off}^{(x)}(t),\hat{L}_i^\dagger]\|\right)
\end{align}
We decompose $\hat{\Delta}$ and $\hat{L}_i$ as done in \cref{sec:LR}: $\hat{\Delta}_i=\sum_{l=0}^\infty\hat{\Delta}_{i,l}$ and $\hat{L}_i=\sum_{l=0}^\infty\hat{K}_i^l$.
$\hat{\Delta}_{i,l}$ and $\hat{K}_i^l$ are acting nontrivially onto the region $R_{i,l}=\{j\in\Lambda: d(i,j)\leq l\}$, and their norms are bounded by \cref{eq:norm_Delta_il} and \cref{eq:K_norm}, respectively.

We now evaluate $\sum_{x,i\in \Lambda}\|[\hat{O}_\mathrm{off}^{(x)}(t),\hat{\Delta}_i]\|$.
It is decomposed as
\begin{align}
&\sum_{x,i\in \Lambda}\|[\hat{O}_\mathrm{off}^{(x)}(t),\hat{\Delta}_i]\|
\nonumber \\
&\leq\sum_{x\in\Lambda}\sum_{l=0}^\infty\sum_{i:d(R_x,i)\leq l}2\|\hat{O}_\mathrm{off}^{(x)}(t)\|\cdot\|\hat{\Delta}_{i,l}\|
\nonumber \\
&\leq\sum_{l=0}^\infty(2\ell_\gamma+2l)\gamma(C+2)e^{-(l-1)/\eta}\sum_{x\in\Lambda}\|\hat{O}_\mathrm{off}^{(x)}(t)\|.
\end{align}
By using \cref{eq:accelerated_decay} and $\ell_\gamma\sim\tilde{\gamma}^{-1/2}$, and dropping non-essential constant factor, we obtain
\begin{equation}\label{eq:Ldagger_O_bound}
\sum_{x,i\in \Lambda}\|[\hat{O}_\mathrm{off}^{(x)}(t),\hat{\Delta}_i]\|
\sim v\tilde{\gamma}^{1/2}\left(\zeta e^{-a\tilde{\gamma}^{1/2}vt}+\zeta'\tilde{\gamma}e^{-a'\gamma t}\right)
\end{equation}
for small $\tilde{\gamma}$.
It is easily confirmed that $\sum_{x,i\in\Lambda}\frac{\sqrt{\gamma}}{4}\|[\hat{O}_\mathrm{off}^{(x)}(t),\hat{L}_i]\|$ and $\sum_{x,i\in\Lambda}\frac{\sqrt{\gamma}}{4}\|[\hat{O}_\mathrm{off}^{(x)}(t),\hat{L}_i^\dagger]\|$ are of the same order.
Thus, by substituting \cref{eq:Ldagger_O_bound} into $\sum_{i\in X(t)}\|\mathcal{L}_i^\dagger\hat{O}_X^\mathrm{d}(t)\|$ in \cref{eq:bar_epsilon_mod2-i}, after integration, we obtain
\begin{equation}
\bar{\epsilon}_\mathrm{md}^\mathrm{(2-i)}=O(\tilde{\gamma}^{1/2}\tildetauB).
\end{equation}

\subsubsection*{(ii) $i\in X^c(t)$ and $d(X,j)\leq d(X,i)$}
Let us start with the expression
\begin{align}\label{eq:bar_ep_mod2-ii}
&\bar{\epsilon}_\mathrm{md}^\mathrm{(2-ii)}\leq\frac{\gamma\tauB}{2}\int_0^\infty dt\sum_{i\in X(t)}\sum_{j:d(X,j)\leq d(X,i)}
\nonumber \\
&\left(\|\hat{O}_X^\mathrm{d}(t),\hat{\Delta}_i]\|+\frac{\sqrt{\gamma}}{4}\|[\hat{O}_X^\mathrm{d}(t),\hat{L}_i]\|+\frac{\sqrt{\gamma}}{4}\|\hat{O}_X^\mathrm{d}(t),\hat{L}_i^\dagger]\|\right).
\end{align}

By putting $l=d(X,i)/p$, we do the following decompositions:
\begin{equation}\label{eq:decomp_Delta_L}
\left\{
\begin{aligned}
&\hat{\Delta}_i=\hat{\Delta}_i^l+(\hat{\Delta}_i-\hat{\Delta}_i^l), \\
&\hat{L}_i=\hat{L}_i^l+(\hat{L}_i-\hat{L}_i^l),
\end{aligned}
\right.
\end{equation}
where
\begin{equation}
\hat{\Delta}_i^l\coloneqq \frac{1}{2i}\int_{-\infty}^\infty ds\int_{-\infty}^\infty ds'\, \mathrm{sgn}(s-s')g_i(s)g_i(-s')\hat{A}_i^l(s)\hat{A}_i^l(s').
\end{equation}
The operators $\hat{L}_i^l$ and $\hat{A}_i^l$ were introduced in \cref{sec:LR}: they are operators localized to the region $R_{i,l}=\{j\in\Lambda: d(i,j)\leq l\}$.

By substituting \cref{eq:decomp_Delta_L} into \cref{eq:bar_ep_mod2-ii}, we obtain
\begin{align}\label{eq:bar_ep_mod2-ii-2}
&\bar{\epsilon}_\mathrm{md}^\mathrm{(2-ii)}\leq\frac{\gamma\tauB}{2}\int_0^\infty dt\sum_{i\in X^c(t)}\sum_{j:d(X,j)\leq d(X,i)}
\nonumber \\
&\left(2\|\hat{\Delta}_i-\hat{\Delta}_i^l\|+\|[\hat{O}_X^\mathrm{d}(t),\hat{\Delta}_i^l]\|+\frac{\sqrt{\gamma}}{2}\|\hat{L}_i-\hat{L}_i^l\|
\right. \nonumber \\
&\left.+\frac{\sqrt{\gamma}}{4}\|[\hat{O}_X^\mathrm{d}(t),\hat{L}_i^l]\|+\frac{\sqrt{\gamma}}{4}\|[\hat{O}_X^\mathrm{d}(t),\hat{L}_i^{l\dagger}]\|\right).
\end{align}

As for $\|\hat{\Delta}_i-\hat{\Delta}_i^l\|$, by using the inequality $\|\hat{A}_i(s)-\hat{A}_i^l(s)\|\leq\min\{2,Ce^{-(l-v|s|)/\xi}\}$, which is obtained from the Lieb-Robinson bound, the operator norm is evaluated as
\begin{align}
\|\hat{\Delta}_i-\hat{\Delta}_i^l\|&\leq \frac{\gamma}{8}(Ce^{-l/2\xi}+2e^{-l/v\tauB})
\nonumber \\
&\leq\frac{\gamma}{8}(C+2)e^{-l/\eta},
\end{align}
where $\eta=\max\{2\xi, v\tauB\}$.
Next, as for $\|[\hat{O}_X^d(t),\hat{\Delta}_i^l]\|$ in \cref{eq:bar_ep_mod2-ii-2}, by using $d(X,R_{i,l})=d(X,i)-l=d(X,i)(p-1)/p$ and $\|\hat{\Delta}_i^l\|\leq \gamma/8$, the Lieb-Robinson bound yields
\begin{equation}
\|[\hat{O}_X^d(t),\hat{\Delta}_i^l]\|\leq \frac{\gamma C}{8}e^{-\left(\frac{p-1}{p}r-vt\right)/\xi},
\end{equation}
where $r=d(X,i)$.
As for $\|\hat{L}_i-\hat{L}_i^l\|$, we can use \cref{eq:jump_localize}:
\begin{equation}
\|\hat{L}_i-\hat{L}_i^l\|\leq\frac{\sqrt{\gamma}}{2}(C+2)e^{-l/\eta}.
\end{equation}
Finally, as for $\|[\hat{O}_X^\mathrm{d}(t),\hat{L}_i^l]\|$ and $\|[\hat{O}_X^\mathrm{d}(t),\hat{L}_i^{l\dagger}]\|$, by using $d(X,R_{i,l})=d(X,i)(p-1)/p$ and $\|\hat{L}_i\|\leq\sqrt{\gamma}/2$, we have
\begin{equation}
\|[\hat{O}_X^\mathrm{d}(t),\hat{L}_i^l]\|=\|[\hat{O}_X^\mathrm{d}(t),\hat{L}_i^{l\dagger}]\|
\leq\frac{\sqrt{\gamma}C}{2}e^{-\left(\frac{p-1}{p}r-vt\right)/\xi}.
\end{equation}

By collecting them, \cref{eq:bar_ep_mod2-ii-2} is evaluated by approximating the sum over $i$ by the integral:
\begin{equation}
\sum_{i\in X^c(t)}\sum_{j:d(X,j)\leq d(X,i)}\approx\int_{pvt}^\infty dr\, (|X|+2r).
\end{equation}
As a result, we find
\begin{equation}
\bar{\epsilon}_\mathrm{md}^\mathrm{(2-ii)}\lesssim \tilde{\gamma}^2\tilde{\tau}_\mathrm{B}
\end{equation}
for small $\gamma$ or small $\tauB$.

\subsubsection*{(iii) $j\in X^c(t)$ and $d(X,j)\leq d(X,i)$}

In \cref{eq:bar_epsilon_mod2}, by using $\hat{O}_{X,j}^\mathrm{d}(t)$, which is defined in \cref{eq:O_Xj}, we have
\begin{align}\label{eq:mod2_norm}
\|[\tilde{\mathcal{L}}_i^\dagger\hat{O}_X^\mathrm{d}(t),\hat{A}_j(\nu s)]\|
\leq \|[\tilde{\mathcal{L}}_i^\dagger\hat{O}_{X,j}^\mathrm{d}(t),\hat{A}_j(\nu s)]\|
\nonumber \\
+2\|\tilde{\mathcal{L}}_i^\dagger(\hat{O}_X^\mathrm{d}(t)-\hat{O}_{X,j}^\mathrm{d}(t))\|.
\end{align}
We now decompose $\tilde{\mathcal{L}}_i$ as
\begin{equation}
\tilde{\mathcal{L}}_i^\dagger=\sum_{l=0}^\infty\tilde{\mathcal{L}}_{i,l}^\dagger
=\sum_{l'=0}^l\tilde{\mathcal{L}}_{i,l'}^\dagger+\sum_{l'=l+1}^\infty\tilde{\mathcal{L}}_{i,l'}^\dagger,
\end{equation}
where $l$ is chosen as 
\begin{equation}\label{eq:mod2_l}
l=\left\lfloor\frac{d(X,j)}{4p}\right\rfloor,
\end{equation}
and $\tilde{\mathcal{L}}_{i,l}^\dagger$ is defined by
\begin{equation}
\tilde{\mathcal{L}}_{i,l}^\dagger\hat{O}=i[\hat{\Delta}_{i,l},\hat{O}]+\mathcal{D}_{i,l}\hat{O}
\end{equation}
for an arbitrary operator $\hat{O}$.
See \cref{eq:D_il,eq:Delta_il} for the definition of $\mathcal{D}_{i,l}$ and $\hat{\Delta}_{i,l}$, respectively.
\Cref{eq:mod2_norm} is then bounded as
\begin{align}\label{eq:mod2_norm2}
\|[\tilde{\mathcal{L}}_i^\dagger\hat{O}_X^\mathrm{d}(t),\hat{A}_j(\nu s)]\|
\leq \left\|\left[\sum_{l'=0}^l\tilde{\mathcal{L}}_{i,l'}^\dagger\hat{O}_{X,j}^\mathrm{d}(t),\hat{A}_j(\nu s)\right]\right\|
\nonumber \\
+2\sum_{l'=l+1}^\infty\|\tilde{\mathcal{L}}_{i,l'}^\dagger\|_\mathrm{cb}+2\|\tilde{\mathcal{L}}_i^\dagger\|_\mathrm{cb}\|\hat{O}_X^\mathrm{d}(t)-\hat{O}_{X,j}^\mathrm{d}(t)\|.
\end{align}
If we choose $l$ as in \cref{eq:mod2_l}, the distance between the support of $\sum_{l'=0}^l\tilde{\mathcal{L}}_{i,l'}^\dagger\hat{O}_{X,j}^\mathrm{d}(t)$ and the site $j$ is greater than $d(X,j)/2p$ (the former is $R_{i,l}\cup X_j$ under the condition $R_{i,l}\cap X_j\neq\emptyset$).
Therefore, the Lieb-Robinson bound yields
\begin{align}
&\left\|\left[\sum_{l'=0}^l\tilde{\mathcal{L}}_{i,l'}^\dagger\hat{O}_{X,j}^\mathrm{d}(t),\hat{A}_j(\nu s)\right]\right\|
\nonumber \\
&\leq\left\|\sum_{l'=0}^l\tilde{\mathcal{L}}_{i,l'}^\dagger\hat{O}_{X,j}^\mathrm{d}(t)\right\|\max\{2,Ce^{-\left(\frac{d(X,j)}{2p}-vs\right)/\xi}\}.
\end{align}
Since
\begin{equation}
\sum_{l'=0}^l\hat{O}=i[\hat{\Delta}_i^l,\hat{O}]+\hat{L}_i^{l\dagger}\hat{O}\hat{L}_i^l-\frac{1}{2}\{\hat{L}_i^{l\dagger}\hat{L}_i^l,\hat{O}\},
\end{equation}
we have
\begin{equation}
\left\|\sum_{l'=0}^l\tilde{\mathcal{L}}_{i,l'}^\dagger\hat{O}_{X,j}^\mathrm{d}(t)\right\|
\leq 2\|\hat{\Delta}_i^l\|+2\|\hat{L}_i^l\|^2\leq \frac{3}{4}\gamma,
\end{equation}
where we have used $\|\hat{\Delta}_i^l\|\leq\gamma/8$ and $\|\hat{L}_i^l\|\leq \sqrt{\gamma}/2$.
By using \cref{eq:norm_D_il,eq:norm_Delta_il,eq:mod2_l}, we find that there is a constant $D_\eta$ that depends solely on $\eta$ and $C$ such that 
\begin{equation}
\sum_{l'=l+1}^\infty\|\tilde{\mathcal{L}}_{i,l'}\|_\mathrm{cb}\leq\gamma D_\eta e^{-\eta d(X,j)/4p}.
\end{equation}
As for $\|\tilde{\mathcal{L}}_i\|_\mathrm{cb}\|\hat{O}_X^\mathrm{d}(t)-\hat{O}_{X,j}^\mathrm{d}(t)\|$ in \cref{eq:mod2_norm2},
\begin{equation}
\|\tilde{\mathcal{L}}_i\|_\mathrm{cb}\leq 2\|\hat{\Delta}_i\|+2\|\hat{L}_i\|^2\leq\frac{3}{4}\gamma
\end{equation}
and the Lieb-Robinson bound implies
\begin{equation}
\|\hat{O}_X^\mathrm{d}(t)-\hat{O}_{X,j}^\mathrm{d}(t)\|\leq C|X|e^{-\left[\frac{p-1}{p}d(X,j)-vt\right]/\xi}.
\end{equation}

By collecting the above results, \cref{eq:mod2_norm2} leads to
\begin{align}
\|[\tilde{\mathcal{L}}_i^\dagger\hat{O}_X^\mathrm{d}(t),\hat{A}_j(\nu s)]\|\leq
\frac{3}{4}\gamma\max\{2,Ce^{-\left(\frac{d(X,j)}{2p}-vs\right)/\xi}\}\nonumber \\
+\gamma D_\eta e^{-\eta d(X,j)/4p}+\frac{3}{2}\gamma C|X|e^{-\left[\frac{p-1}{p}d(X,j)-vt\right]/\xi}.
\end{align}
By substituting it into \cref{eq:bar_epsilon_mod2} with the condition $i\in X^c(t)$ and $d(X,j)\leq d(X,i)$, we can explicitly calculate an upper bound on $\bar{\epsilon}_\mathrm{md}^\mathrm{(2-iii)}$.
We do not give lengthy calculations here, but it turns out that the upper bound is finite and 
\begin{equation}
\bar{\epsilon}_\mathrm{md}^\mathrm{(2-iii)}\lesssim\tilde{\gamma}^2\tilde{\tau}_\mathrm{B}
\end{equation}
for small $\gamma$ or small $\tauB$.

\bibliography{apsrevcontrol,physics}

\end{document}